\documentclass[12pt]{iopart}
\expandafter\let\csname equation*\endcsname\relax
\expandafter\let\csname endequation*\endcsname\relax

\usepackage{etoolbox}

\makeatletter
\def\@mkboth#1#2{}
\newlength\appendixwidth
\preto\appendix{\addtocontents{toc}{\protect\patchl@section}}
\newcommand{\patchl@section}{%
  \settowidth{\appendixwidth}{\textbf{Appendix }}%
  \addtolength{\appendixwidth}{1.5em}%
  \patchcmd{\l@section}{1.5em}{\appendixwidth}{}{\ddt}%
}
\makeatother

\usepackage{cite}
\usepackage[colorlinks, citecolor = cyan, urlcolor = blue]{hyperref}

\usepackage{epsfig}
\usepackage{latexsym}
\usepackage{bm}

\usepackage{graphicx}

\usepackage{amsmath}
\usepackage{amsfonts}
\usepackage{amssymb}
\usepackage[mathscr]{eucal}

\usepackage{xcolor}

\usepackage{color}
\definecolor{dgreen}{rgb}{0,0.7,0}

\def\el{\ell}
\begin{document}

\title{Non-local linear response in anomalous transport}
\author{Anupam Kundu}
\address{International Centre for Theoretical Sciences, Tata Institute of Fundamental Research, Bengaluru 560089, India}
\ead{anupam.kundu@icts.res.in}
\date{\today}

\begin{abstract}
\noindent
{Anomalous heat transport observed in low dimensional classical systems is associated to super-diffusive spreading of space-time correlation of the conserved fields in the system. This leads to  non-local linear response relation between the heat current and the local temperature gradient in non-equilibrium steady state. This relation provides a generalisation of Fourier's law of heat transfer and is characterised by a non-local kernel operator which is related to fractional operators describing super-diffusion. The kernel is essentially proportional, in appropriate hydrodynamic scaling limit, to  the time integral of the space-time correlations of local currents in equilibrium. In finite size systems, the time integral of correlation of microscopic currents at different locations over infinite duration is independent of the locations. On the other hand the kernel operator is space-dependent. We demonstrate that the resolution of this apparent puzzle appears through taking appropriate combination of limits of large system size  and large integration time duration. Our study shows the importance of taking the limits in proper way even for (open) systems connected to reservoirs. In particular we reveal how to extract the kernel operator from simulation data of microscopic current-current correlation. For two model systems exhibiting anomalous transport, we provide direct and detailed numerical verification of the kernel operators.}
\end{abstract}

%\maketitle
\makeatletter
\tableofcontents

\section{Introduction} 
Transport of heat through materials plays the role of a paradigm in non-equilibrium statistical physics. In macroscopic systems this phenomena is commonly described by the Fourier's law. However, many numerical as well as analytical studies suggest that the Fourier's law is  not valid in one dimensional momentum conserving systems \cite{dhar2008heat,lepri2016thermal,kundu2019review}. This fact is usually demonstrated by the system size $N$ scaling of  the heat conductivity 
%$\kappa$ as  
\begin{align}
\kappa \sim N^\nu,
\end{align}
where the exponent $\nu$ characterizes the nature of transport being ballistic for $\nu=1$, diffusive for $\nu=0$ and anomalous for $0 < \nu <1$. In the anomalous and the ballistic cases the thermal conductivity diverges with increasing system size. 

For systems exhibiting anomalous transport, the Fourier's law gets manifested as a non-local linear response (LR) relation characterised by a kernel $\mathcal{K}(u,v)$ operator that replaces the finite conductivity \cite{kundu2019review,kundu2019fractional,kundu2018anomalous,cividini2017temperature,miron2019derivation}. The kernel operator is   related to  the time integral of the space-time correlations of local hydrodynamic currents \cite{kundu2019review,kundu2019fractional,kundu2022super,kundu2018anomalous,cividini2017temperature,miron2019derivation}. In this paper, we however show that such time integral  of the space-time correlations of local currents in system with finite number of degrees of freedom becomes independent of the locations where the currents are measured. On the other hand the kernel operator is location dependent. This generates an apparent puzzle. In this paper we resolve this puzzle and disclose how to extract the information about the kernel from the space-time correlation of the microscopic currents, as measured in simulations. In particular, we provide direct numerical verifications of the detailed analytical forms of the kernel operators in two microscopic models that exhibit anomalous transport.

%\noindent

%\section{The problem and it's relevance}
A commonly followed approach for determining the $N$ dependence of thermal conductivity $\kappa$ is to use the Green-Kubo (GK) formula relating thermal conductivity to the time integral of the total heat current auto-correlation function in equilibrium. 
 There are two types of GK formulae: (a) Isolated system GK formula and (b) open system GK formula \cite{purkayastha2019classifying}. In the former case,  the GK formula reads
\begin{align}
\kappa =  \lim_{\tau \to \infty}  \lim_{N \to \infty}  \frac{1}{k_BT^2N^d} \int_0^\tau dt \langle J(t) J(0)\rangle_{eq}, \label{ISGK}
\end{align}
where, $J(t)$ is the total current in an isolated system of linear dimension $N$  in $d$-dimension at temperature $T$ and $\langle ... \rangle_{eq}$ represents average in equilibrium.   The order of the limits in Eq.~\eqref{ISGK} is important and in the correct order, the right hand side relates to the response of an isolated system to some external drive {\it e.g.} temperature difference or chemical potential bias. There are several derivations of the GK formula in Eq.~\eqref{ISGK} \cite{green1954markoff,kubo1957statistical,mori1958statistical,green1960comment,luttinger1964theory,visscher1974transport}. 
While the formula in Eq.~\eqref{ISGK} is important, it can not be straightforwardly used in many low dimensional systems which, as mentioned earlier, may exhibit anomalous transport {\it i.e.} diverging thermal conductivity $\kappa$. In order to obtain the nature of divergence in such cases the usual procedure that has been followed is to put a cut-off at $t_c \sim N$ in the the upper limit of the time integral in Eq.~\eqref{ISGK}. Although there is no rigorous justification for assuming such a cut-off, however one considers it reasonable in systems having sound modes. 
 
 Another way to obtain a GK formula  in large but finite size system is to consider the open system set-up in which the system is connected to reservoirs of temperatures  $T_L=T+\Delta T/2$ and $T_R=T-\Delta T/2$ at the two ends. For small temperature difference $\Delta T$ between the reservoirs, exact linear response (LR) relations similar to the GK formula in Eq.~\eqref{ISGK} have been derived \cite{kundu2009green,gallavotti1996extension,lebowitz1999gallavotti, rey2001fluctuations,purkayastha2019classifying}. For a one dimensional system of size $N$ defined on a lattice (for example, chain of particles with nearest neighbour interaction) the average of the local energy current $\langle j^{(e)}_{i,i-1} \rangle_{\Delta T}$ in the non-equilibrium steady state (NESS) from $(i-1)^{\rm th}$ site (particle) to $i^{\rm th}$ site (particle) can be expressed in terms of the time integral of the current-current correlation in equilibrium as 
 \begin{align}
\langle j^{(e)}_{i+1,i} \rangle_{\Delta T} &= -\sum_{m=1}^{N-1}K_N(i,m)~(T_{m+1}-T_m),
\label{LRR-1}
\end{align}
for small $\Delta T$ where 
% \begin{align}
%K_N(i,m)&=\frac{1}{T^2} \int_0^\infty dt \langle j^{(e)}_{i+1,i}(t)j^{(e)}_{m+1,m}(0) \rangle_{eq},
%\label{Klm-1}
%\end{align}
 \begin{align}
K_N(i,m)&=\int _0^\infty dt~ \mathscr{C}_N(i,m,t),~~\text{where},~
 \mathscr{C}_N(i,m,t)=\frac{1}{T^2}\langle j^{(e)}_{i+1,i}(t)j^{(e)}_{m+1,m}(0)\rangle_{eq}.
\label{Klm-1}
\end{align}
and $T_m$ represents the local temperature at the $m-$th site in the NESS. In this paper we have set $k_B=1$.
The equation \eqref{LRR-1} is also a  LR relation. By computing the time integral of current-current correlation $K_L(i,m)$ in equilibrium and finding the system size dependence of the sum, one can estimate the system size scaling of the stationary current in the NESS and hence the value of the exponent $\nu$. Usually both the isolated system GK formula and the open system GK formula provide the same value for the exponent, however may not  always do so \cite{das2014heat,deutsch2003correlations,purkayastha2018anomalous}. There are only  few systems for which the current-current correlation or the time integral of it can be computed analytically \cite{kundu2010time,morgado2009exact}, but for systems with generic interaction potential it is difficult. For such systems one requires to rely on numerical simulations.

In the last decade there has been a significant development in the hydrodynamic description of anomalous transport in one-dimensional systems which is known as non-linear fluctuating hydrodynamic (NFHD) theory\cite{spohn2014nonlinear,spohn2016fluctuating,das2014numerical}. In this theory 
%\bluew{one takes a zoomed out view of the particles on a mesoscopic length scale (which is much larger than the microscopic scale but much smaller than the macroscopis scale). At this scale the particles appear as a continuous medium like a fluid and  the continuity equations for the conserved quantities  are now replaced by hydrodynamic (HD) equations however one still has fluctuations.} 
the local conserved quantities, say energy $e_i$ now gets replaced by a density field $e(x,t)$ and the corresponding HD current  $j^{(e)}(x,t)$ satisfy the continuity equations which give rise to the HD equations under slowly varying and slowly evolving local equilibrium assumption. The dissipations and the noise terms in the HD equations are added phenomenologically obeying fluctuation-dissipation relation. Often the dissipation at the mesoscopic scales are modelled by diffusion terms. One writes similar HD equations for other conserved quantities as well, such as stretch and momentum.

The NFHD theory makes precise predictions for the scaling forms of the space-time correlation of the density fields of the conserved quantities. Depending on the inter couplings of the conserved fields at the nonlinear order in deviations  from the underlying global equilibrium values, the NFHD theory classifies the transport properties being diffusive, super diffusive and ballistic for a wide class of systems both Hamiltonian and stochastic \cite{spohn2016fluctuating, spohn2014nonlinear, spohn2015nonlinear, mendl2014equilibrium, kundu2019review, miron2019derivation, cividini2017temperature, popkov2016exact, popkov2015fibonacci,popkov2014superdiffusive,chakraborty2019dynamics}. Using specific sum rules and the scaling forms for the space-time correlations for the density fields, one can make predictions for the decay (more precisely the exponent of the power law decay) of the correlations of the corresponding HD currents with time\cite{spohn2016fluctuating,spohn2014nonlinear,mendl2015current}. Using this correlation in the isolated system GK formula one can find the exponent $\nu$ for a  closed system in the macroscopic limit.

In the context of open system set-up the NFHD formalism has been used recently in two model systems called harmonic chain with momentum exchange (HCME) and harmonic chain with volume exchange (HCVE) to compute the hydrodynamic current-current correlations \cite{cividini2017temperature, kundu2022super}. In these systems one finds that the above LR relation in Eq.~\eqref{LRR-1} gets modified to
\begin{align}
\langle j^{(e)}(x) \rangle_{\Delta T} = -\int_0^N {K}_N^{\rm hd}\left(x,y \right)  \frac{d T}{dy}dy, \label{LRR-hd}
\end{align}
%\begin{align}
%\langle j^{(e)}(x) \rangle_{\Delta T} = -\int_0^N K_N(x,y) \frac{d T}{dy}dy, \label{LRR-hd}
%\end{align}
where ${K}_N^{\rm hd}(x,y) = \int _0^\infty dt~ \mathscr{C}_N^{\rm hd}(x,y,t),$
and $T(x)$ is the  temperature profile in the NESS which connects the temperatures $T_L$ and $T_R$ at the left and right ends of the system respectively. The superscript `hd' in $\mathscr{C}_N^{\rm hd}(x,y,t)$ represents the space-time correlation  hydrodynamic currents at different locations computed using fluctuating hydrodynamic equations.

It has been shown analytically as well as numerically that both the HCME and HCVE models exhibit anomalous transport with exponent $\nu =1/2$ {\it i.e.} the current $J_{ss} = \langle j_e(x) \rangle_{\Delta T}$ in the NESS decays with system size as $J_{ss} = \frac{\mathscr{J}_{ss}\Delta T}{\sqrt{N}}$ for large $N$ \cite{lepri2008stochastic, olla2009thermal, basile2006momentum, kundu2018anomalous, kundu2019review} (where $\mathscr{J}_{ss}$ is a constant) 
%\be
%J_{ss} \sim \frac{\mathscr{J}\Delta T}{\sqrt{N}}, \label{J_ss-vs-N}
%\end{equation}
and the temperature profile $T(x)$ possesses the scaling form $T(x) = T + \Delta T~\theta\left(\frac{x}{N} \right)$
%\be
%T(x) = T + \Delta T~\theta\left(\frac{x}{N} \right),~\label{T(x)-sf}
%\end{equation}
such that $\theta(0) = 1/2$ and $\theta(1)=-1/2$. Inserting these large $N$ scaling forms for the stationary current $J_{ss}$ and the temperature $T(x)$ in Eq.~\eqref{LRR-hd}  one can rewrite the LR relation as 
\begin{align}
\mathscr{J}_{ss}=-\int_0^1 dv~\mathcal{K}(u,v)~\partial_v\theta(v). \label{LRR-sc}
\end{align}
where the kernel $\mathcal{K}(u,v)$ is obtained from
%\begin{align}
%\mathcal{K}(u,v)=\lim \limits_{c \to \infty} \lim \limits_{N\to \infty} \sqrt{N} \int_0^{cN} dt~\mathscr{C}_N(uN,vN,t),
%\label{kernel-scaling-lim}
%\end{align}
%\frac{|u-v|N}{c}
\begin{align}
\mathcal{K}(u,v)=\lim \limits_{N \to \infty} \sqrt{N} \int_0^{\infty} dt~\mathscr{C}_N^{\rm hd}(uN,vN,t).
\label{kernel-scaling-lim}
\end{align}
It has been shown analytically that the limit in the above equation indeed exists for both the HCME and HCVE model and one finds explicit forms for the kernel $\mathcal{K}(u,v)$ \cite{cividini2017temperature, kundu2022super, kundu2019review, kundu2022super}. Note the equation \eqref{LRR-sc} generalises the usual Fourier's law (which is a local LR relation) to a non-local LR relation. In fact for the choice $\mathcal{K} \propto \delta(u-v)$, which occurs for the  diffusive transport, the LR equation  \eqref{LRR-sc} reduces to the usual Fourier's law. As in the diffusive case, solving the Fourier law equation $\mathscr{J}=-\kappa~\partial_v\theta(v)$, one can find the temperature profile $\theta(v)$. Similarly, one can solve the  Eq.~\eqref{LRR-sc} to obtain $\theta(v)$ in the anomalous transport case. 
%For the HCME and HCVE models this equation has been solved to compute the temperature profiles which are verified using direct numerical simulations.
%Indeed it has been done for the HCME and HCVE model and verified against numerical simulation, which indirectly established the existence of the limit in Eq.~\eqref{scaling-lim} as well as the detailed form of the kernel $\mathcal{K}(u,v)$.

In this paper we present an argument based on calculations involving microscopic currents, which implies that the  kernel $K_N(i,m)$ involving microscopic currents defined in Eq.~\eqref{Klm-1} can not depend on locations $i$ and $m$  where the currents are measured. On the other hand, the kernel $\mathcal{K}(u,v)$ involving hydrodynamic currents is space dependent. Indirect numerical verification for the existence of the limit in Eq.~\eqref{kernel-scaling-lim} as well as the detailed analytical forms of the kernel $\mathcal{K}(u,v)$ have been obtained for both HCME and HCVE model by comparing the solution $\theta(v)$ of Eq.~\eqref{LRR-sc} with the same obtained from numerical simulation. This generates an apparent puzzle which we resolve in this paper. More precisely we address the following precise questions:
%\item by computing the current-current correlation in equilibrium and estimating the system size scaling of the time integral one can can obtain the system size scaling of the current.
%\item In few specific cases the current-current correlation can be computed analytically --- my harmonic chain paper, Archak's paper etc.
%\item generic interaction potential it is difficult to compute this correlation analytically and one needs to revert to numerical simulation
%\item however using NFHD it has  been possible to compute similar correlations among the corresponding hydrodynamic currents for a class of interacting systems in the isolated system setting....cite Herbert's paper
%\item In the context of open system set-up the NFHD formalism has been used very recently to compute the hydrodynamic current-current correlations. In this context the above LR relation gets modified to .....
%\item this relation provides a generalisation to Fourier's law ..... our review
%\item In particular for HCME and HCVE model one finds explicit expressions of the kernel which are then used to compute the temperature profiles.
%\item However, an argument based on microscopic calculation shows $K(i,m)$ can not depend on both $i$ and $m$ and it is consistent with other derivation of GK relation in open system.
%\item on the other hand the hydrodynamic calculations provide a space dependent kernel which are indirectly verified by comparing the temperature profiles computed using the kernel and the same obtained from simulation.
\begin{itemize}
\item How to reconcile the LR response formula in Eq.~\eqref{LRR-1} and Eq.~\eqref{LRR-sc}? 
\item How to understand the HD LR formula in Eq.~\eqref{LRR-sc} from a simulation point of view in which one 
is often forced to work with the microscopic currents?
\end{itemize}
We support our answers with extensive numerical evidence for the HCME and HCVE model systems. 

The paper is organised as follows. In section \ref{model-def} we describe the two model systems HCME and HCVE  microscopically. We also discuss the allowed conservations laws and the associated currents. Next we discuss the form of the global equilibrium distributions in this section. In the next section \ref{Deri-LRR} we provide a linear-response theory based on estimating correction to local equilibrium state and show how one can possibly get a non-local linear response relation, where non-local nature appears through a kernel which is related to time integral of the current-current correlation. Such correlations are difficult to compute from microscopic calculations. However, sometimes it is easier to compute such correlations using fluctuating hydrodynamic theory which for the two models are presented separately in Sec.~\ref{FHD}. In this section we show that the Fourier's law for diffusive transport gets replaced by a non-local linear response relation characterized by a non-local kernel operator that is related to time integral of the current-current correlations. However, in Sec.~\ref{puzzle} we argue that for a finite size system, such time integral over infinite time duration can not yields a space-dependent kernel -- giving rise to an apparent puzzle. In Sec.~\ref{resolution} we resolve this puzzle and present extensive numerical results which not-only provides direct evidence of the existence of the space-dependent kernel but also offers numerical verification of the detailed analytical forms of the kernel operators. In the last section \ref{conclusion} we provide conclusions. Some of the calculations are relegated to the appendices.

\section{Descriptions of the models and conservation laws}
\label{model-def}
In this section we describe the HCME and HCVE model. Both models are defined on one-dimensional lattice of size $N$. Each lattice site contains one particle of unit mass which carries some degrees of freedom. The particles at the left and right ends are connected to reservoirs of temperatures $T_L=T+\frac{\Delta T}{2}$ and $T_R=T-\frac{\Delta T}{2}$, respectively, which we model as Langevin baths. Particles in the bulk evolve according to some deterministic equations.  In order to have good ergodic properties in the system,   the particles are additionally subjected to bulk noises that keeps all the conservation laws of the deterministic evolution valid. In generic real  systems one has non-linear interactions which are believed to provide the mechanisms of necessary ergodicity properties such that the system reaches stationary states locally quickly. Since performing analytical computations with more realistic interaction potentials is difficult, one often takes a complementary approach by introducing stochastic part in the dynamics which has the same conservation laws as the non-linear dynamical system and also makes the system ergodic \cite{lepri2016thermal,kundu2019review,miron2019derivation,lepri2008stochastic,lepri2011density,lepri2010nonequilibrium}. 
Stochastic models such as HCME and HCVE have played an important role in understanding the transport of energy or more generally other conserved quantities allowed by the dynamics \cite{lepri2008stochastic,lepri2011density,bernardin2012anomalous,spohn2015nonlinear,miron2019derivation}.
%This is mainly due to the fact that the stochastic approach seems to easily yield results that would require much more efforts by adopting the dynamical approach. On the other hand the properties at macroscopic scale such as hydrodynamic evolution does not crucially depend on the details of the microscopic details of the Hamiltonian except for some non-universal constants (such as transport coefficients).
Below we provide the details of the dynamics  for the two models separately.

\subsection{Harmonic chain with momentum exchange: } The particles at each lattice site carries a position (or displacement) and a momentum variable. Let $q_i$ and $p_i$ represent the position and momentum of the $i-$th particle. The particles are interacting via harmonic interaction of strength $k_{harm}=\omega^2>0$. The boundary particles are subject to different boundary conditions (BCs).  In addition the momenta of the neighbouring sites are exchanged randomly with a constant rate $\gamma$. 
The equations of motion for $i=1,2,...,N$ are 
%\begin{align}
%\begin{split}
%\frac{dq_i}{dt}&=p_i \\
%\frac{dp_i}{dt} &= (1-\delta_{i,1}-\delta_{i,N})~k (q_{i+1}-2 q_i+q_{i-1})\\ 
%& +\delta_{i,1}\left[ k(q_2 -\zeta q_1) -\lambda p_1 +\xi_1\right] \\
%& + \delta_{i,1}\left[ k(q_{N-1} -\zeta q_N) -\lambda p_N +\xi_N\right] \\
%&+ \text{exchange of}~p_i~\text{with}~p_{i-1}~\text{or}~p_{i+1}~\text{with rate}~\gamma,
%\end{split}
%\end{align}
\begin{align}
\begin{split}
\frac{dq_i}{dt}&=p_i \\
\frac{dp_i}{dt} &= {\omega^2} (q_{i+1}-2 q_i+q_{i-1})
+\delta_{i,1}\left(-\lambda p_1 +\sqrt{2\lambda T_L}\xi_1\right) 
+ \delta_{i,1}\left( -\lambda p_N +\sqrt{2\lambda T_R}\xi_N\right), \\
&~~~~~~~~~~~+~ \text{exchange of}~p_i~\text{with}~p_{i-1}~\text{or}~p_{i+1}~\text{at rate}~\gamma, 
\end{split}
\label{hcme-leq}
\end{align}
where
\begin{eqnarray}
q_0=q_1,~&q_{N+1}=q_N,~&~\text{for free BC}, \label{hcme-freeBC}\\
q_0=0,~&q_{N+1}=0,~&~\text{for fixed BC}. \label{hcme-fixBC}
\end{eqnarray}
%$q_0=q_1,~q_{N+1}=q_N$ and for fixed boundary condition (BC) $q_0=q_{N+1}=0$. 
The noises $\xi_1$ and $\xi_N$ are two 
white Gaussian noises of zero mean and unit variance and $\lambda$ is the strength of the dissipation. For $T_L=T_R=T$ the system starting from an arbitrary configuration reaches, after a long time,  an equilibrium state described by the Gibbs distribution
\begin{align}
P_{eq}(\{q_i,p_i\}) &= \frac{e^{-\frac{H}{T}}}{\mathcal{Z}_{m}},~\text{with}~
H(\{q_i,p_i\}) = \sum_{i=1}^N e_i, \label{hcme-gibbs}\\ 
\text{where,}~~s_i&= q_{i+1}-q_i,~\text{and}~e_i = \frac{p_i^2}{2} + \frac{\omega^2}{4}(s_i^2+s_{i-1}^2). \label{loc-s-e}
%\sum_{i=1}^{N}\frac{p_i^2}{2} + \frac{1}{2} \sum_{i=1}^N \left[(q_{i}-q_{i-1})^2\right],
\end{align}
Here $\delta_{i,j}$ is Kronecker delta  and $\mathcal{Z}_m$ is the partition function. For $T_L \neq T_R$ the system reaches a NESS with currents following across  the system. In order to identify the currents, we look at the conserved quantities. For the HCME model without the baths at the ends, the total stretch $\sum_{i=0}^{N}s_i$, total momenta $\sum_{i=1}^{N}p_i$ and total  energy $\sum_{i=0}^N e_i$ remain conserved. These conservation laws imply  continuity equations for the local stretch $s_i$, local momenta $p_i$ and local energy $e_i$. The continuity equations read
\begin{align}
\frac{do_i}{dt} = j^{(o)}_{i,i-1}-j^{(o)}_{i+1,i},~~~\text{with},~~o=(s,p,e),~~\text{for}~~i=1,2,...,N, \label{hcme-conti-eq}
\end{align}
where $j^{(o)}_{i,i-1}$ is the energy current coming from the $(i-1)^{\rm th}$ site to the $i^{\rm th}$ site corresponding to the locally conserved quantity $o_i$. The explicit expressions of the currents for $i=1,...,N$ are 
\begin{align}
\begin{split}
j^{(s)}_{i+1,i} &= -p_{i+1}, \\
j^{(p)}_{i+1,i} &= -{\omega^2}s_i + \sigma_{i+1,i}(p_i-p_{i+1})+ \delta_{i,0}\left(-\lambda q_1+\sqrt{2\lambda T_L} \xi_1\right) 
- \delta_{i,N}\left( -\lambda q_N+\sqrt{2\lambda T_R} \xi_N\right) \\ 
%\text{and} &   \\
j^{(e)}_{i+1,i} &=-\frac{{\omega^2}}{2}(p_i+p_{i+1})s_i + \sigma_{i+1,i}\left(\frac{p_i^2}{2}-\frac{p_{i+1}^2}{2}\right) \\ 
& ~~~~~~~+\delta_{i,0}p_1(-\lambda q_1 +\sqrt{2\lambda T_L}\xi_1)
- \delta_{i,N}p_N(-\lambda q_N +\sqrt{2\lambda T_R}\xi_N).
\end{split}
\label{hcme-micro-j}
\end{align}
where $\sigma_{i+1,i}$ for $i=1,2,...,N-1$ represents the exchange process occurring between $i^{\text{th}}$ and $(i+1)^{\text{th}}$ particles, which are independent Poisson processes with rate $\gamma$. 

\subsection{Harmonic chain with  volume exchange:} 
Now we describe the HCVE model. In this case the particles on each lattice site carry a single variable $\eta_i \in \mathbb{R}$ which in the literature is called the `volume' variable \cite{bernardin2012anomalous,spohn2015nonlinear,kundu2018anomalous}. However, to make notations consistent across the two the models, we in this paper imagine it as a `displacement' variable. There is some local energy $V(\eta_i)$ associated to each lattice site given by  $V(\eta_i)=k_{\rm o}\frac{\eta_i^2}{2}$ with $k_{\rm o}>0$. The variable $\eta_i$ evolves deterministically under the influence of this local energy but from the neighbouring sites. In addition, as in the HCME model, the displacement variables from the neighbouring sites are exchanged at random with rate $\gamma$. The first and the $N$-th site are attached to two Langevin reservoirs of temperatures $T_L$ and $T_R$, respectively. The evolution of equations  for $i=1,2,...,N$ are given by 
\begin{align}
\frac{d \eta_i}{dt}&=k_{\rm o}(\eta_{i+1}-\eta_{i-1}) + \delta_{i,1}(-\lambda k_{\rm o} \eta_1 + \sqrt{2\lambda T_L}\xi_1) 
+ \delta_{i,N}(-\lambda k_{\rm o} \eta_N + \sqrt{2\lambda T_R}\xi_N)  \\
&~~~~~~~~~~~+ ~\text{exchange of}~\eta_i~\text{with}~\eta_{i-1}~\text{or}~\eta_{i+1}~\text{at rate}~\gamma, 
\end{align}
with BCs $\eta_0=0$ and $\eta_{N+1}=0$. For equal temperatures of the reservoirs, this model in the thermodynamic limit reaches an invariant state described by the Gibbs distribution \cite{bernardin2012anomalous}
\begin{align}
P_{eq}(\{\eta_i\}) = \prod_{i=1}^N \sqrt{\frac{k_{\rm o}}{2 \pi T}}e^{-\frac{k_{\rm o}}{2T}\eta_i^2}.
\end{align}
In this state the average displacement and energy per particle are $\langle \eta_i \rangle_{eq}=0$ and $\langle V(\eta_i)\rangle_{eq}$ $=k_{\rm o}T/2$. 

In this model one has two locally conserved quantities ---  local displacement $h_i=\eta_i$ and local energy $e_i=V(\eta_i)=k_{\rm o}\eta_i^2/2$ which satisfy continuity equations in Eq.~\eqref{hcme-conti-eq} with the following expressions for the currents
\begin{align}
\begin{split}
j^{(h)}_{i+1,i}&= -k_{\rm o}(\eta_i+\eta_{i+1}) + \sigma_{i+1,i}(\eta_i-\eta_{i+1})  \\ 
& ~~~~~~~+ \delta_{i,1}(-\lambda k_{\rm o} \eta_1 + \sqrt{2\lambda T_L}\xi_1) - \delta_{i,N}(-\lambda k_{\rm o} \eta_N + \sqrt{2\lambda T_R}\xi_N) \\
j^{(e)}_{i+1,i}&= -k_{\rm o}^2 \eta_i\eta_{i+1} + \sigma_{i+1,i}k_{\rm o}\left(\frac{\eta_i^2}{2}-\frac{\eta_{i+1}^2}{2}\right) \\ 
& ~~~~~~~+ \delta_{i,1}k_{\rm o}\eta_1(-\lambda k_{\rm o} \eta_1 + \sqrt{2\lambda T_L}\xi_1) - \delta_{i,N}k_{\rm o}\eta_N(-\lambda k_{\rm o} \eta_N + \sqrt{2\lambda T_R}\xi_N).
\end{split}
\label{hcve-micro-j}
\end{align}
When $T_L\neq T_R$ the system reaches a NESS in which one finds a non-zero energy current but zero displacement current because the reservoirs do not provide a `pressure' difference across the system. For small $\Delta T$, the stationary energy current can be related to the time integral of the local current-current correlation in the LR regime. We provide a brief derivation of this LR relation in the next section.

\section{Derivation of the LR  in Eq.~\eqref{LRR-1}}
\label{Deri-LRR}
Let us discuss the derivation of the LR Eq.~\eqref{LRR-1} for the HCME model and for this we follow the procedure given in \cite{kundu2022super}. We start with the Fokker-Planck (FP) equation for the joint distribution $P(\vec{\mu},t)$  with $\vec{\mu}=\{q_i,p_i\}$,
\begin{equation}
\partial_t P(\vec{\mu},t) = \mathcal{L} P(\vec{\mu},t),~~\text{where}~\mathcal{L} = \mathcal{L}_\ell + \mathcal{L}_{ex} + \mathcal{L}_b. \label{FP-hcme}
\end{equation}
Here $\mathcal{L}_\ell$ represents the Liouvilian part, $\mathcal{L}_{ex}$ represents the exchange part of the FP operator $\mathcal{L}$. The operator $\mathcal{L}_b$ contains the contribution from the reservoirs at the boundaries. Explicit expressions of these operators are provided in \ref{app1}. 

Since the HCME system has good ergodic properties, the system locally reaches a local equilibrium (LE) state at a much smaller time scale than it takes to reach the global equilibrium state or the NESS depending on whether $\Delta T$ is zero or not.  We assume that the system starts in a local equilibrium state given by
\begin{align}
P_{\rm le}(\vec{\mu}) =\frac{1}{Z_{\rm le}(\{T_i,\pi_i,\tau_i\})}\prod_{i=1}^Ne^{-(\frac{e_i}{T_i} +\pi_ip_i +\tau_i s_i) },
\label{P_le-hcme}
\end{align}
where $Z_{\rm le}$ is the normalisation constant. The distribution $P_{\rm le}$ is 
characterised by the local temperature, pressure and momentum profiles $T_i=T_i(0),\tau_i=\tau_i(0)$ and $\pi_i=\pi_i(0)$ at time $t=0$ which are slowly varying over space. 
 As the system evolves, this LE state also evolves slowly in time because of the conservation laws and is characterised by space-time dependent  fields $\{T_i(t), \pi_i(t), \tau_i(t) \}$. Time evolutions of these fields can be obtained by averaging both sides of the continuity equations \eqref{hcme-conti-eq} with respect to $P(\vec{\mu},t)$, which essentially provide the macroscopic HD evolutions of these fields. Since we focus only on thermal drive across boundaries, sensible boundary conditions for the LE fields are $\pi_0(t)=\pi_{N+1}(t)=0$, $\tau_0(t)=\tau_{N+1}(t)=0$, $T_0(t)=T_L$ and $T_{N+1}(t)=T_R$.

In the linear response regime {\it i.e.} for small $\Delta T$,  the actual joint distribution $P(\{q_i,p_i\},t)$ remains always close to the time evolved LE state. It is reasonable to  write the solution of the FP equation \eqref{FP-hcme} at a later time $t$ as
\begin{equation}
P(\vec{\mu},t) = P_{\rm le}(\vec{\mu},t) + P_d(\vec{\mu},t), \label{P-tot}
\end{equation}
where the distribution $P_d(\vec{\mu},t)$ represents the deviation from the LE distribution which satisfies 
\begin{equation}
\partial_t P_d(t) -\mathcal{L} P_d(t) = \mathcal{L}P_{\rm le}(t) - \partial_t P_{\rm le}(t),
\end{equation}
with $P_d(\vec{\mu},0)=0$. 
Formal solution of this equation is given by
\begin{align}
P_d(\vec{\mu},t) = \int_0^t dt' e^{\mathcal{L}(t-t')} \left[ \Phi(\vec{\mu},t) - \Phi_{le}(\vec{\mu},t)\right] P_{\rm le}(\vec{\mu},t),
\end{align}
where 
\begin{equation}
\Phi(\vec{\mu},t) = \frac{\mathcal{L}P_{\rm le}(\vec{\mu},t)}{P_{\rm le}(\vec{\mu},t)},~~\text{and}~~
\Phi_{le}(\vec{\mu},t) =\frac{\partial_t P_{\rm le}(\vec{\mu},t)}{P_{\rm le}(\vec{\mu},t)}. \label{Phi}
\end{equation}
Explicit expressions of $\Phi$ and $\Phi_{le}$ are provided in  \ref{app2}. 
Using the form for the full distribution $P(\vec{\mu},t)$ from Eq.\eqref{P-tot}, one can compute the average local currents 
$\langle j^{(a)}_{i+1,i}(t) \rangle_{P=P_{\rm le}+P_d} $ where the microscopic currents $j^{(a)}_{i+1,i}$ are given in 
 Eq.~\eqref{hcme-micro-j}. In order to get the average  local currents in the NESS one takes the $t \to \infty$ limit and get
 \begin{align}
 \langle j^{(a)}_{i+1,i}(t) \rangle_{P} &=  \langle j^{(a)}_{i+1,i} \rangle_{le} + \int d\vec{\mu}  \int_0^\infty dt' j^{(a)}_{i+1,i} \left( e^{\mathcal{L}(t-t')} \left[ \Phi - \Phi_{le}\right] P_{\rm le}(\vec{\mu},t)\right), \\
 &= \langle j^{(a)}_{i+1,i} \rangle_{le} +  \int_0^\infty dt' \left \langle j^{(a)}_{i+1,i}(t) \left[ \Phi(\vec{\mu},t) - \Phi_{le}(\vec{\mu},t)\right] P_{\rm le}(\vec{\mu},t)\right \rangle _{le},  \label{av-j-1}
 \end{align}
where $\langle ... \rangle_{le}$ represents average over the LE distribution. For small $\Delta T$ in the LR regime, the average $\langle ... \rangle_{le}$ can be replaced by average over global equilibrium state [Eq.~\eqref{hcme-gibbs}] because $[\Phi -\Phi_{le}]$ is already of order $\Delta T$. Furthermore,  in the LR regime the first part $\langle j^{(a)}_{i+1,i}(t) \rangle_{le}$ in Eq.~\eqref{av-j-1} gets contribution only from the exchange events which are in the gradient form ( {\it i.e.} decays as $\sim 1/N$) and cannot provide the expected anomalous contribution (which decays as $\sim 1/\sqrt{N}$). Hence we neglect this term. Now inserting the explicit forms of 
$[\Phi -\Phi_{le}]$ from Eq.~\eqref{app:Phi-Phi_le} and keeping only the leading order terms one gets the LR relation in Eq.~\eqref{LRR-1}.
% (see Appendix \ref{app3} for details).
%\be
%\langle j^{(e)}_{i+1,i} \rangle_{\Delta T} = -\frac{1}{T^2} \sum_{m=1}^{N-1}K_N(i,m)~(T_{m+1}-T_m), \label{LRR-2}
%\end{equation} 
%as announced in Eq.~\eqref{LRR-1}. 

A similar calculation has been carried out for the HCVE model in \cite{kundu2022super}. In this case also one gets the LR relation  in Eq.~\eqref{LRR-1} with only difference now is that one should use the instantaneous currents from Eq.~\eqref{hcve-micro-j}.

\section{Fluctuating hydrodynamics and generalised Fourier's law}
\label{FHD}
Often it is difficult to compute the kernel $K_N(i,m)$ from microscopic calculations. In such situations, fluctuating HD theory  provides a way to compute various space-time correlations of the densities as well as the associated currents \cite{spohn2014nonlinear,spohn2016fluctuating,das2014numerical}. In the HD theory one takes a zoomed out view of the particles on a mesoscopic length scale (which is much larger than the microscopic scale but much smaller than the macroscopis scale). At this scale the particles appear as a continuous medium like a fluid and  the continuity equations for the conserved quantities $o_i(t)$ are now replaced by hydrodynamic (HD) equations for the corresponding field densities represented by $o(x,t)$ in the continuum limit,  however they still contain fluctuations. 

%cividini2017temperature,
\subsection{HCME:} Following the prescription of the NFHD framework \cite{spohn2014nonlinear, spohn2015nonlinear, mendl2014equilibrium,   popkov2016exact}, we write the equations for the HCME model in terms of the sound modes $\phi_{\pm}(x,t)=\omega s(x,t) \mp p(x,t)$ and the heat mode $\phi_0(x,t) $ $= e(x,t)$ \cite{cividini2017temperature} as 
\begin{align}
\partial_t \phi_{\pm}(x,t) &= - \partial_s \left[ \pm \omega \phi_\pm(x,t) - D \partial_x \phi_\pm(x,t) - \sqrt{2 D} \zeta_\pm(x,t)\right], \label{eq:phi_pm} \\
\partial_t \phi_0(x,t) &= - \partial_s \left[ \frac{\omega}{4}\left(\phi_+^2(x,t) -\phi_-^2(x,t) \right)  - D_0 \partial_x \phi_0(x,t) - \sqrt{2D_0} \zeta_0(x,t)\right], 
\label{eq:phi_0}
\end{align}
where the diffusion and noise terms are added phenomenologically. The noises $\zeta_\pm(x,t)$ and $\zeta_0(x,t)$ are white Gaussion noise with zero mean and delta correlation both in space and time. The instantaneous energy current density $j^{(e)}(x,t)$ can be easily read from Eq.~\eqref{eq:phi_0}
\begin{equation}
j^{(e)}(x,t) = \frac{\omega}{4}\left(\phi_+^2(x,t) -\phi_-^2(x,t) \right), \label{J_e-hd-hcme}
\end{equation}
where we have neglected the subdominant contributions from the diffusion terms and noises. In terms of this current, one can write a continuum form of the the LR relation 
%in Eq.~\eqref{LRR-1} in the continuum limit now reads 
as given in Eq.~\eqref{LRR-hd}.
% which, for convenience, we here rewrite   
%\begin{align}
%\langle j^{(e)}(x) \rangle_{\Delta T} = -\int_0^N K_N(x,y) \frac{d T}{dy}dy, \label{LRR-hd-2}
%\end{align}
%with
%\begin{align}
%K_N(x,y) = \frac{1}{T^2} \int_0^\infty dt \langle j_e(x,t) j_e(y,0) \rangle_{eq}. \label{kernel-hd}
%\end{align}
The fluctuating field equations for $\phi_\pm(x,t)$ were solved with both fixed and free boundary conditions in \cite{cividini2017temperature}. These solutions provide the explicit expressions of the local currents $j^{(e)}(x,t)$ using which one can compute the correlation $\langle j^{(e)}(x,t) j^{(e)}(y,0) \rangle_{eq}$.  Performing the time integral  one  finds the kernel in Eq.~\eqref{LRR-hd}.  Taking the large $N$ limit one obtaines expressions of the scaled kernel $\mathcal{K}(u,v)$ defined in Eq.~\eqref{kernel-scaling-lim}. The explicit form of the scaled kernel $\mathcal{K}(u,v)$ depends on the boundary condition characterised by an effective reflection coefficient 
\cite{cividini2017temperature, lepri2011density}
\begin{equation}
R = \left(\frac{\lambda -\omega}{\lambda + \omega} \right)^2, \label{refl-coeff}
\end{equation}
for free BC. 
The value  $R=0$ corresponds to pure (no reflection) free BC and $0<R < 1$ corresponds to general free boundaries at which energy gets partially reflected back and the rest gets absorbed  to the reservoirs. For fixed BC $R=1$ irrespective of the values of $\lambda$ and $\omega$. 
For general $R$, the scaled kernel $\mathcal{K}(u,v)$ has the form \cite{cividini2017temperature, kundu2019review}
\begin{align}
\mathcal{K}_{hcme}(u,v) = \frac{\mathscr{A}}{\sqrt{2\pi}} \left[ \sum_{{n=-\infty}}^\infty \left( \frac{R^{|2n|}}{\sqrt{|2n+u-v|}} -  \frac{R^{|2n+1|}}{\sqrt{|2n+u+v|}}\right)\right],  \label{kernel-hcme}
\end{align}
%\begin{align}
%\mathcal{K}_{hcme}(u,v) = \frac{\mathscr{A}}{\sqrt{2\pi}} \left[ \frac{1}{\sqrt{|u-v|}} - \frac{R}{\sqrt{|u+v|}}+ \sum_{\underset{n\neq 0}{n=-\infty}}^\infty \left( \frac{R^{|2n|}}{\sqrt{|2n+u-v|}} -  \frac{R^{|2n+1|}}{\sqrt{|2n+u+v|}}\right)\right],  \label{kernel-hcme}
%\end{align}
with $\mathscr{A}=\frac{\omega^{3/2}}{2 \sqrt{2\gamma}}$ \cite{kundu2019review}. Note for the purely free BC ({\it i.e.} $R=0$) the expression of the kernel becomes particularly simple. The expression for the kernel in Eq.~\eqref{kernel-hcme} can be used  in Eq.~\eqref{LRR-sc} to solve for the temperature profiles. Profiles obtained through such procedure were verified with the same obtained from numerical simulation as well as  analytical expressions (whenever possible) obtained through spectral decomposition of fractional Laplacian in bounded domain. Analytical solution for the temperature profiles using spectral method is possible only for the fixed BC case {\it i.e.} for $R=1$ \cite{kundu2019review,cividini2017temperature, lepri2008stochastic}. This provides a indirect validation of the existence of the limit in Eq.~\eqref{kernel-scaling-lim} as well as the non-locality. In this paper we provide a direct numerical verification of the existence of $\mathcal{K}(u,v)$ as well as demonstrate how to obtain it's non-local structure from the correlation of microscopic currents.

\subsection{HCVE:} For the HCVE model, there are two conserved quantities -- displacement and energy. Under assumption of slowly varying local equilibrium picture, the corresponding continuity equations in the continuum limit provides the HD equations in which, once again, the diffusion and noise terms are added phenomenologically. For a detailed discussion on the fluctuating hydrodynamics of this model see \cite{spohn2015nonlinear, kundu2022super}. This model has one sound mode $\phi_h(x,t) = -\sqrt{\frac{k_{\rm o}}{T}}h(x,t)$ and one heat mode $\phi_e(x,t) = -\sqrt{\frac{2}{T}} \left(e(x,t) -\frac{T}{2}\right)$ (since $\langle \eta_i \rangle_{eq} =0$ for our case) \cite{kundu2022super}. In terms of these modes the NFHD equations are written as 
\begin{align}
\partial_t \phi_h(x,t) &= -\partial_x \left[ - 2k_{\rm o} \phi_h(x,t) -D_h \partial_x \phi_h(x,t) - \sqrt{2D_h} \zeta_h(x,t) \right] \label{hcve-sound} \\
\partial_t \phi_e(x,t) &= -\partial_x \left[ - \sqrt{2}k_{\rm o} \phi_h^2(x,t) -D_e \partial_x \phi_e(x,t) - \sqrt{2D_e} \zeta_h(x,t) \right], \label{hcve-heat}
\end{align}
where $D_h,~D_e$ are phenomenological diffusion constants and $\zeta_{h,e}(x,t)$ are white Gaussian noise with zero mean and delta function correlation both in space and time. From Eq.~\eqref{hcve-heat}, one can easily identify the heat current 
\begin{equation}
j^{(e)}(x,t) = -\sqrt{2}k_{\rm o} \phi_h^2(x,t), \label{hd-j-hcve}
\end{equation}
where once again we have neglected the subdominant contributions from the diffusion and the noise terms.
Now computing the space-time current-current correlation in equilibrium and performing the time integral one finds that in the large $N$ limit the scaling form of the kernel $\mathcal{K}(u,v)$ in this model reads as [see \cite{kundu2022super} for derivation]
\begin{align}
\mathcal{K}_{hcve}(u,v) = \frac{k_{\rm o}^{3/2}}{2\sqrt{\pi \gamma}} ~\frac{\Theta(v-u)}{\sqrt{v-u}}, \label{kernel-hcve}
\end{align}
where $\Theta(z)$ is Heaviside theta function.
This form of the kernel has also been derived analytically from a microscopic calculation in \cite{kundu2018anomalous} and  it's form was verified, once again indirectly, by computing  the temperature profile analytically and numerically. The temperature profile in the NESS  has the form $\theta(u) = \sqrt{1-u}-1/2$.   For this model also we provide direct numerical validation of the form of the kernel in Eq.~\eqref{kernel-hcve} from correlations of the microscopic currents.

\begin{figure}
\begin{center}
\leavevmode
\includegraphics[scale=0.4]{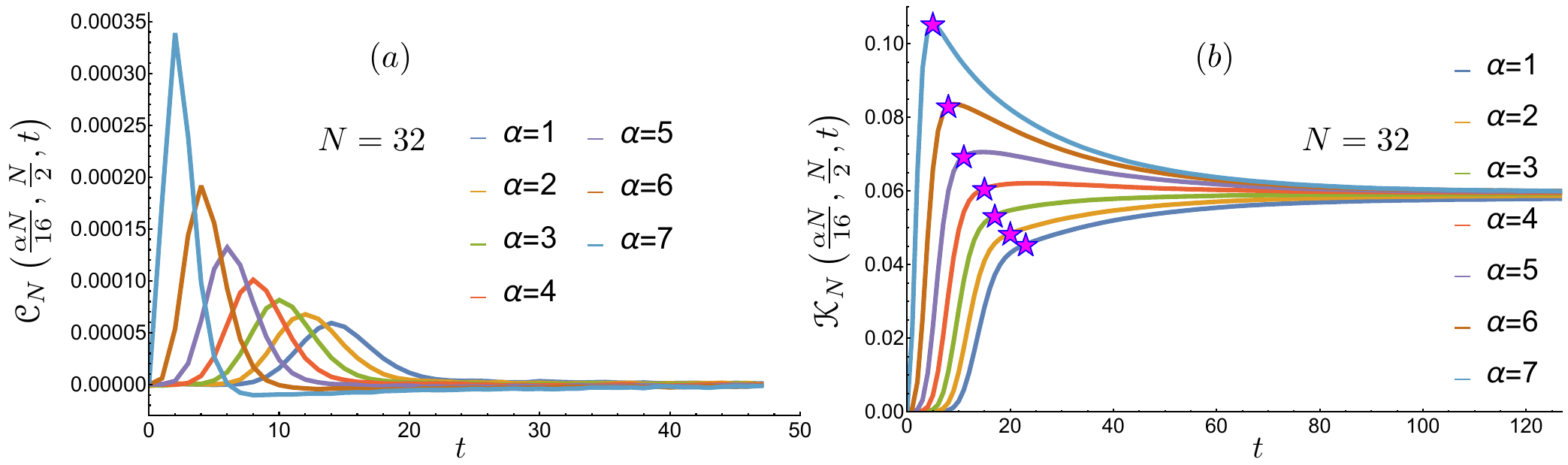}
%	\put (-170,120) {$\textbf{{\small(a)}}$}
%\includegraphics[width=7cm,angle=0]{figs/H-function-diff-dv-N50000_Scipost.eps}
%	\put (-170,120) {$\textbf{{\small(b)}}$}
\caption{Plots showing time evolution of $\mathscr{C}_N(i,m,t)$ and $\mathscr{K}_N(i,m,t)$ for $m = \frac{N}{2}$ and $i=\frac{\alpha N}{16}$ with $\alpha=1,2,...,7$. Parameters for this plot are $\lambda=1,~\omega=1$. }
\label{C-K-R=0}
\end{center}
\end{figure}

\section{An apparent puzzle}
\label{puzzle}
In the previous section we have discussed the existence of the space dependent (scaled) kernel $\mathcal{K}(u,v)$ which is essentially the  large $N$ scaling limit of the time integral of the local (hydrodynamic) current-current correlation in equilibrium. However for a finite size system, we below provide an argument which shows that the kernel $K_N(i,m)$ defined in Eq.~\eqref{Klm-1} can not be dependent on space coordinates $i,m$ on the lattice. To show this we define the quantity 
\begin{align}
Q_i(t) = \int_0^t d t' j^{(e)}_{i+1,i}(t'), \label{int-cur-i}
\end{align}
which measures the net amount of energy current following across the bond $(i,i+1)$. If $\Delta E_{i,m}(t)$ represent the energy between the sites $(i+1)$ and $m$ at time $t$ such that 
\begin{equation}
 \Delta E_{i,m}=\sum_{j=i+1}^m e_j(t)
 \end{equation}
  then it is easy to see
\begin{align}
\Delta E_{i,m}(t) -\Delta E_{i,m}(0) = Q_i(t) - Q_m(t).
\end{align}
We consider the fluctuation of the left hand side of the above equation
\begin{align}
\langle (\Delta E_{i,m}(t) -\Delta E_{i,m}(0))^2 \rangle_{eq} = \langle Q_i^2 \rangle_{eq} + \langle Q_m^2 \rangle_{eq} - 2 \langle Q_i(t)Q_m(t) \rangle_{eq}
\label{E_Q-fluc-r}
\end{align}
Using the time translational and time reversal invariance of the two time correlations in thermal equilibrium, one can show that, in the large $t$ limit, one can rewrite the terms in the above equation as follows 
\begin{align}
\lim \limits_{t \to \infty}\frac{\langle Q_i(t)Q_m(t) \rangle_{eq}}{2tT^2} 
%&= \lim \limits_{t \to \infty} \frac{1}{2t}\int_0^t d t_1 \int_0^t d t_2  \langle j^{(e)}_{\el,\el+1}(t_1)j^{(e)}_{m+1,m}(t_2) \rangle_{eq}, \\
%&= \lim \limits_{t \to \infty} \frac{1}{2t}\int_0^t d t_1\left( \int_0^{t_1}+ \int_{t_1}^t \right)d t_2  \langle j^{(e)}_{\el,\el+1}(t_1)j^{(e)}_{m+1,m}(t_2) \rangle_{eq}, \\
%&= \lim \limits_{t \to \infty} \frac{1}{2t} \left( \int_0^t d t_1 \int_0^{t_1} d t_2  \langle j^{(e)}_{\el,\el+1}(t_1)j^{(e)}_{m+1,m}(t_2) \rangle_{eq} \right. \\
%&~~~~~~~~~~~~~~~~\left. +  \int_0^t d t_2 \int_0^{t_2} d t_1  \langle j^{(e)}_{\el,\el+1}(t_2)j^{(e)}_{m+1,m}(t_1) \rangle_{eq} \right)\\
%&= \lim \limits_{t \to \infty} \frac{1}{t} \int_0^t d t_1 \int_0^{t_1} d t_2  \langle j^{(e)}_{\el,\el+1}(t_1)j^{(e)}_{m+1,m}(t_2) \rangle_{eq}  \\
%&=\int_0^\infty d t_1 \left( 1-\frac{t_1}{t}\right)  \langle j^{(e)}_{\el,\el+1}(t_1)j^{(e)}_{m+1,m}(0) \rangle_{eq}, \\
&= \frac{1}{T^2}\int_0^\infty d t_1  \langle j^{(e)}_{i+1,i}(t_1)j^{(e)}_{m+1,m}(0) \rangle_{eq} = K_N(i,m). 
\label{def:K_n(i,m)}
\end{align}
So for large $t$, 
\begin{align}
\langle Q_i^2 \rangle_{eq} &\simeq 2T^2K_N(i,i)~t, \\ 
\langle Q_m^2 \rangle_{eq} &\simeq 2T^2K_N(m,m)~t, ~~~\text{and}\\ 
\langle Q_i(t)Q_m(t) \rangle_{eq} &\simeq 2T^2K_N(i,m)~t, \label{Qsq-asymp}
\end{align}
in the leading order. Since the system is homogeneous in equilibrium, we should expect that 
\begin{equation}
\lim \limits_{t \to \infty}\frac{\langle Q_\el(t)^2 \rangle_{eq}}{2tT^2} = \lim \limits_{t \to \infty}\frac{\langle Q_m(t)^2 \rangle_{eq}}{2tT^2}.
\end{equation}
This implies $K_N(i,i)=K_N(m,m)=C_N$, where $C_N$ is a $N$ dependent constant but independent of $i$ and $m$. 
On the other hand $\langle (\Delta E_{i,m}(t) -\Delta E_{i,m}(0))^2 \rangle_{eq} $ can not grow with time in equilibrium and for large $t$ it approaches a constant which may depend on $i$ and $m$. Hence, from Eqs.~\eqref{E_Q-fluc-r} and \eqref{Qsq-asymp} we must cancel the linearly growing terms. This implies 
\begin{equation}
K_N(i,m)= C_N, \label{finite-N-kernel}
\end{equation} 
independent of $i$ and $m$. Note that Eq.~\eqref{finite-N-kernel} is valid for for arbitrary but fixed $N$, only $t \to \infty$ limit has been taken on the left hand side [see the definition in Eq.~\eqref{def:K_n(i,m)}]. The GK formula in  Eq.~\eqref{LRR-1} now becomes $\langle j^{(e)}_{i+1,i} \rangle_{\Delta T} = \Delta TC_N$
%\begin{align}
%\langle j^{(e)}_{i+1,i} \rangle_{\Delta T} = \Delta TC_N, \label{GK-finite}
%\end{align}
%\begin{align}
%\langle j^{(e)}_{i+1,i} \rangle_{\Delta T} = \frac{\Delta T}{K_B T^2N} \int_0^\infty \langle J(t)J(0) \rangle_{eq},
%\end{align}
which is consistent with the GK formula derived in \cite{kundu2009green}. One can provide a similar argument for other conserved currents as well.

Now question is: How to reconcile the two facts  that $K_N(i,m)=C_N$ {\it i.e.} independent of $i$ and $m$ and the scaled kernel $\mathcal{K}(u,v)$ is space-dependent. 
To resolve the puzzle one needs to be careful while taking the large system size and large integration time duration limits.  In the next section we show that one requires to take these two limits in a combined fashion {\it i.e.} one requires to integrate first up to a time $\tau_N$ that depends on $N$ and then take the $N \to \infty$ limit. This is similar to what is done  in Eq.~\eqref{ISGK} where 
one takes $\tau_N \sim c N$.

\begin{figure}
\begin{center}
\leavevmode
\includegraphics[scale=0.3]{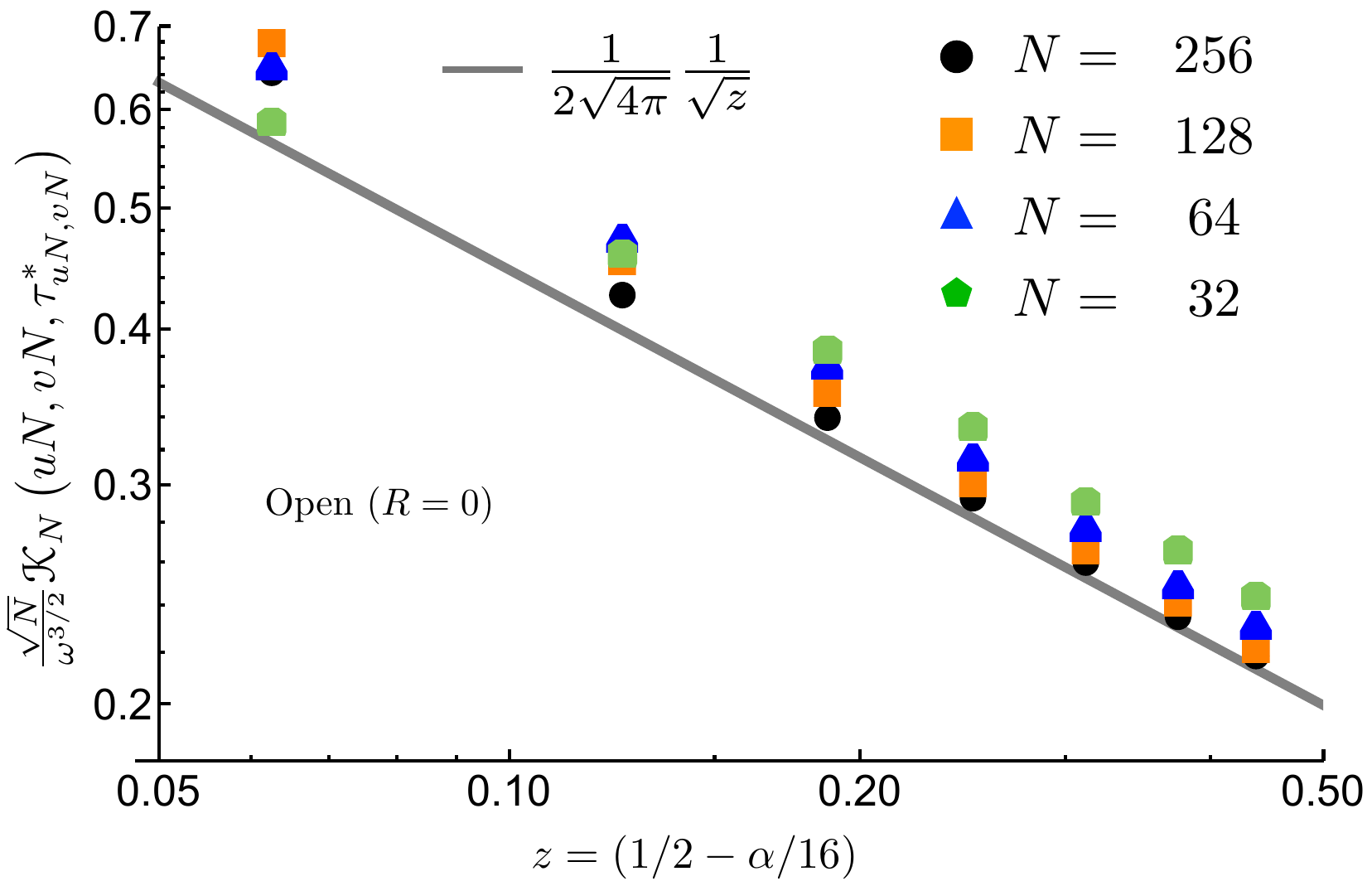}
%	\put (-170,120) {$\textbf{{\small(a)}}$}
%\includegraphics[width=7cm,angle=0]{figs/H-function-diff-dv-N50000_Scipost.eps}
%	\put (-170,120) {$\textbf{{\small(b)}}$}
\caption{Plots of $\frac{\sqrt{N}}{\omega^{3/2}}\mathscr{K}_N\left(uN,vN,t^\star_{uN,vN}\right)$ versus $z=v-u$ with $v =\frac{1}{2}$ and $u=\frac{\alpha}{16}$ for $\alpha=1,2,...,7$. The solid gray line represents the right hand side (RHS) of Eq.~\eqref{mcalK_R-0}. The symbols corresponding to different $N$ are obtained from simulation. Parameters for this plot are same as fig.~\ref{C-K-R=0}. }
\label{num-verify-K-R=0}
\end{center}
\end{figure}

\section{Resolution and numerical support}
\label{resolution}
To resolve the puzzle, we start with the following quantities
\begin{align}
\mathscr{K}_N\left(i,m,t \right) &= \int_0^t dt'~ \mathscr{C}_N(i,m,t'),~\text{with}, \label{K_N(imt)} \\
% \mathscr{C}_N(i,m,t)&=\frac{1}{2T^2}\left[\langle j^{(e)}_{i+1,i}(\tau)j^{(e)}_{m+1,m}(0)\rangle_{eq}+\langle j^{(e)}_{m+1,m}(\tau)j^{(e)}_{i+1,i}(0)\rangle_{eq} \right]. 
 \mathscr{C}_N(i,m,t)&=\frac{1}{T^2}\langle j^{(e)}_{i+1,i}(t)j^{(e)}_{m+1,m}(0)\rangle_{eq}. 
 \label{C_N(imt)}
\end{align}
 The integrand $ \mathscr{C}_N(i,m,\tau)$ in Eq.~\eqref{K_N(imt)} represents the equilibrium time correlation between energy currents at the $i^{\rm th}$ and $m^{\rm th}$ bonds. For given $N$, the quantity $\mathscr{K}_N(i,m,\tau \to \infty) =$ $ K_N(i,m)=C_N$ with $C_N$ being proportional to the the stationary current in the system and hence independent of $i$ and $m$. On the other hand the following limit provides  
 \begin{align}
 \lim_{N\to \infty} \sqrt{N}\mathscr{K}_N\left(uN,vN,\frac{|u-v|N}{c}\right) = \mathcal{K}(u,v),  \label{def:mcalK}
 \end{align}
 for both the models HCME and HCVE, where $c$ is speed of the sound modes. To numerically demonstrate this we show that for large $N$, the quantity $\mathscr{K}_N\left(i,m,\tau \right)$ satisfies the following scaling form
%\begin{align}
%\lim_{N\to \infty} \sqrt{N}\mathscr{K}_N\left(uN,vN,t_n(u,v)N+g\sqrt{2 \gamma t_n(u,v)N}\right) = \mathcal{K}_n(u,v), \label{def:mcalK_n}
%\end{align}
\begin{align}
\lim_{N\to \infty} \sqrt{N}\mathscr{K}_N\left(uN,vN, t_{n}(u,v;N)+g\frac{\sqrt{2 \gamma t_{n}(u,v;N)}}{c}\right) = \mathcal{K}_n(u,v), \label{def:mcalK_n}
\end{align}
with some scaling function $\mathcal{K}_n(u,v)$ where $g>0$ is some constant and  $t_{n}(u,v;N)$ is  the time required for a sound peak with speed $c$ to reach site $i=uN$ starting from site $m=vN$ after making $n$ reflections with the boundaries at $0~(u=0)$ and $N~(u=1)$. As the sound mode moves ballistically it also spreads diffusively. The spread of the sound mode at time $t$ is $\sqrt{2 \gamma t}$ ( where $\gamma$ is the diffusion constant of the sound mode \cite{kundu2019review,kundu2022super}). Hence the time required for the sound mode to pass through the site $i=uN$ completely (after making $n$ reflections with the boundaries) is approximately given by  $t_{n}(u,v;N) + g\frac{\sqrt{t_{n}(u,v;N)}}{c}$. One should choose the value of $g$ appropriately such that at a given location there is no overlap between the passing of a sound peak and the arrival of the next sound peak. In our simulation we have chosen  $g= 3/2$ for the HCME model and $g=5/2$ for the HCVE model to account for the diffusive spreading of the sound modes. 

The speed $c$ of the sound mode in the HCME model is $c=\omega$ and in the HCVE model $c=2 k_{\rm o}$. As the number $n$ of reflections increases, the function $\mathcal{K}_n(u,v)$ approaches the full kernel $\mathcal{K}(u,v)$ {\it i.e.} $\mathcal{K}(u,v)=\lim_{n \to \infty}\mathcal{K}_n(u,v)$.  For some boundary conditions, the sound mode does not reflect back at all from the boundaries. In that case $n=0$ and one finds $\mathcal{K}(u,v)=\mathcal{K}_0(u,v)$. 

In the next we present numerical data supporting the above approach. We first present our numerical results  for the HCME model  and then we discuss the same for the HCVE model. For all our simulations we have used integration time step $dt=0.01$ and correlations are obtained averaging over $10^9$ realizations. Also we have chosen $\gamma=1$ for all our simulations and $T=1$ for the HCME model and $T=3$ for the HCVE model.
% while values of other parameters  $N$, $c$ and $\lambda$ are changed as required in different contexts and are mentioned in the discussions of respective figures. 
%\section{Numerical support}

\begin{figure}
\begin{center}
\leavevmode
\includegraphics[scale=0.4]{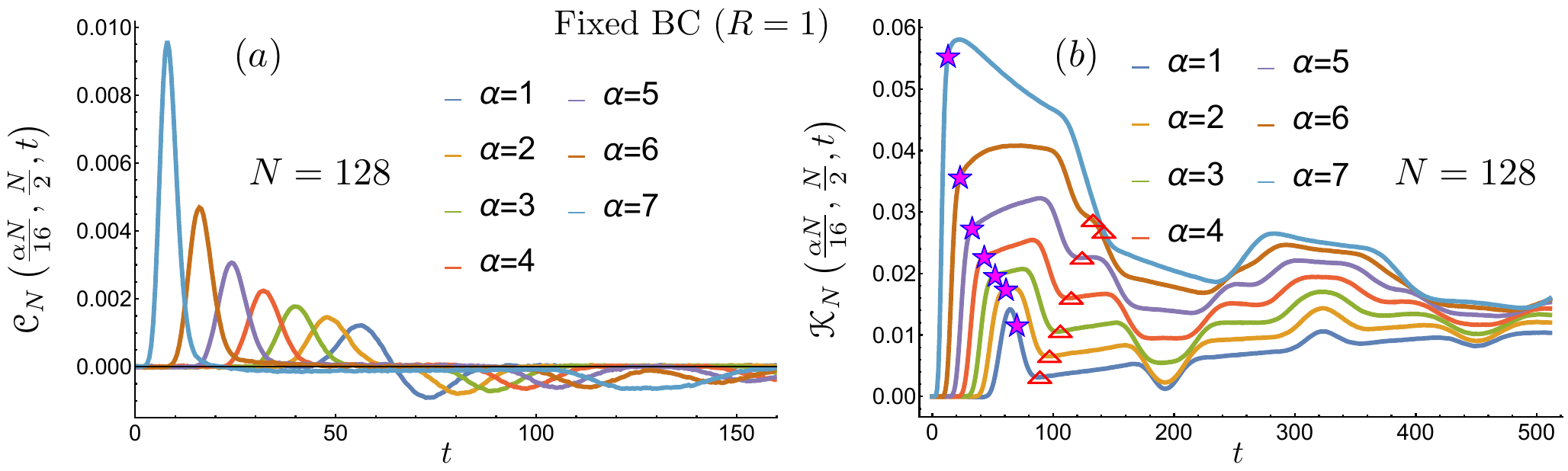}
%	\put (-170,120) {$\textbf{{\small(a)}}$}
%\includegraphics[width=7cm,angle=0]{figs/H-function-diff-dv-N50000_Scipost.eps}
%	\put (-170,120) {$\textbf{{\small(b)}}$}
\caption{Plots showing time evolution of $\mathscr{C}_N(i,m,t)$ and $\mathscr{K}_N(i,m,t)$ for $R=1$ with $m = \frac{N}{2}$ and $i=\frac{\alpha N}{16}$ where $\alpha=1,2,...,7$. Parameters for this plot are $\lambda=1,~\omega=1$.  }
\label{C-K-R=1}
\end{center}
\end{figure}

\subsection{Results for HCME model:}We numerically integrate the Langevin equations \eqref{hcme-leq} and compute space-time correlation $\mathscr{C}_N(i,m,t)$ and its time integration $\mathscr{K}_N(i,m,t)$ defined in Eq.~\eqref{C_N(imt)} 
%for $m = \frac{N}{2}$ and $i=\frac{kN}{16}$ for $k=1,2,...,7$, and 
for different values of $\lambda$, $\omega$ and $N$.  We discuss the purely absorbing (free BC, $R=0$), partially absorbing (partially free BC, $0<R< 1$) and fixed (reflecting BC, $R=1$) BC cases separately.

\subsubsection{$R=0$ case:} This is the pure absorbing boundary case in which $\lambda=\omega$. We choose $\omega=1$, hence the sound speed $c=1$. 
The sound modes in this case get fully absorbed (at the reservoirs) when they reach the boundaries. In fig.~\ref{C-K-R=0}a and \ref{C-K-R=0}b, we plot $\mathscr{C}_N(i,m,t)$ and $\mathscr{K}_N(i,m,t)$, respectively, as functions of time $t$ for $N=32$. Different curves correspond to $i=\frac{\alpha N}{16}$ with $\alpha=1,2,...,7$ and for fixed $m = \frac{N}{2}$. The peaks in fig.~\ref{C-K-R=0}a for different $\alpha$ correspond to the arrival of a sound mode to the position $i$ starting from $m$ and the time required for that is $t_{im}= \frac{|i-m|}{c}$. At this time the spread of sound mode due to diffusion is given by the standard deviation $\sigma_{im} = \sqrt{2 \gamma t_{im}}$. Consequently, the cumulative correlation $\mathscr{K}_N(i,m,t)$ starts increasing significantly from zero  at around $t^{\circ}_{im} = t_{im} -g\frac{\sigma_{im}}{c}$ and reaches the highest value at around  $t^\star_{im} = t_{im} +g\frac{\sigma_{im}}{c}$ (the $\star$ points in fig.~\ref{C-K-R=0}b), after  which it's rate of increase starts decreasing and it finally saturates to a value independent of $u$ and $v$. The fact that  the time integrated correlations for different $i$  saturate to the same value is in fact the numerical verification of Eq.~\eqref{finite-N-kernel} and the saturation value is actually the conductance according to Eq.~\eqref{LRR-1}.

In order to verify the scaling in Eq.~\eqref{def:mcalK_n}, we collect the values of the cumulative correlation $\mathscr{K}_N(i,m,t^\star_{im})$ scaled with $\sqrt{N}$ ({\it i.e.} multiplied with $\sqrt{N}$) at times $t^\star$ for $m=N/2$ and different $i= \alpha N/16$ with $\alpha =1,2,...,7$. The values of the scaled cumulative correlation $\frac{\sqrt{N}}{\omega^{3/2}}\mathscr{K}_N(i,m,t^\star_{im})$ at these time instances should provide the kernel [as can be seen from Eq.~\eqref{kernel-hcme} with $R=0$]
\begin{equation}
\mathcal{K}(u,v) = \frac{\omega^{3/2}}{2\sqrt{4 \pi}} \frac{1}{\sqrt{|u-v|}}, \label{mcalK_R-0}
\end{equation}
when plotted as function of $z=\frac{m}{N}-\frac{i}{N}$.  We verify this expression in fig.~\ref{num-verify-K-R=0} where we plot $\frac{\sqrt{N}}{\omega^{3/2}}\mathscr{K}_N\left(uN,vN,t^\star_{uN,vN}\right)$ against $z=v-u$ in log-log scale for different $N$. We observe data for different $N$ falls on  straight lines and the lines converge towards the theory line (gray solid line) with increasing $N$ which verifies the analytical form of the Kernel in Eq~\eqref{mcalK_R-0}.

\begin{figure}
\begin{center}
\leavevmode
\includegraphics[scale=0.4]{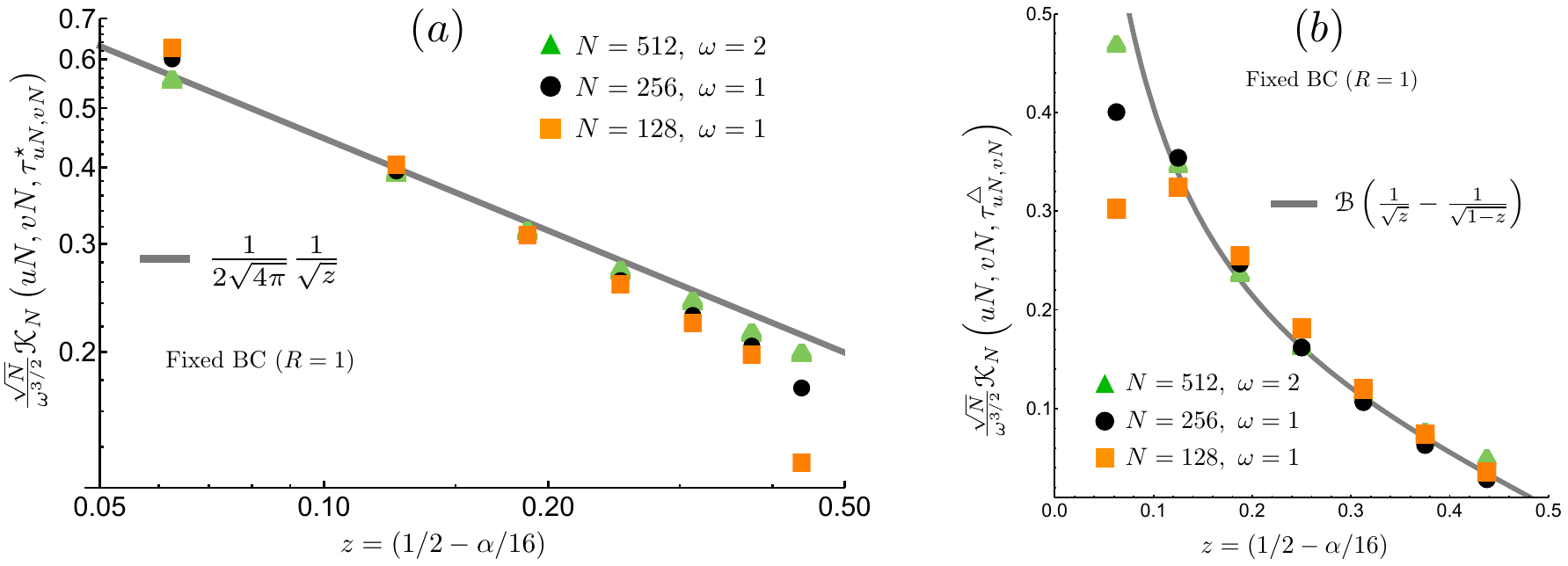}
%	\put (-150,120) {$\textbf{{\small(a)}}$}
%\includegraphics[scale=0.25]{fig/Ker-verify-R-1-2nd.pdf}	
%\includegraphics[width=7cm,angle=0]{figs/H-function-diff-dv-N50000_Scipost.eps}
%	\put (-30,120) {$\textbf{{\small(b)}}$}
\caption{Plots of (a) $\frac{\sqrt{N}}{\omega^{3/2}}\mathscr{K}_N\left(uN,vN,t^\star_{uN,vN}\right)$ {\it vs.} $z=v-u$  (b) $\frac{\sqrt{N}}{\omega^{3/2}}\mathscr{K}_N\left(uN,vN,t^\triangle_{uN,vN}\right)$ {\it vs.} $z$ with $v =\frac{1}{2}$ and $u=\frac{\alpha}{16}$ for $\alpha=1,2,...,7$. The solid grey lines in (a) and in (b) represents the kernel till the first term  and the kernel till the second term on the right hand side (RHS) of Eq.~\eqref{mcalK_R-1}, respectively. The symbols corresponding to different $N$ are obtained from simulation. We have used $\lambda=1$ and $\omega=1$ for this plot. The constant $\mathscr{B}$ in the expression of the theoretical curve [see legend of (b)] is fitted to value $0.192$ whereas the value of the actual constant is $\frac{1}{2\sqrt{4 \pi}}=0.141$. The difference arises possibly due to either finite size effect or not so small numerical integration step $dt$ or both. }
\label{num-verify-K-R=1-12}
\end{center}
\end{figure}

%$\triangle$

\subsubsection{$R=1$ case:} In this case the sound modes moving with speed $c=\omega$ get completely reflected from the boundaries. As a result, the two sound modes starting from some location, say $m=N/2$, crosses a given location $i$ multiple times each after every reflections from the boundaries. Hence, one observes multiple peaks in the plot of $\mathscr{C}_N(i,m,t)$ {\it vs.} $t$ in fig.~\ref{C-K-R=1}a, each corresponding to  passing of a sound mode through the location $i$. After each reflection, the sound mode density profile changes phase (sign) as well as direction  \cite{cividini2017temperature}. A right moving mode gets converted to a left moving mode after reflection at the boundary. When the initial excitation is created at  the middle of the system, the two sound modes $\phi_\pm$ get reflected from the right and left boundaries at the same time and become $- \phi_\mp$. In this situation the instantaneous current at a location  in equilibrium, changes sign after each reflection [see Eq.~\eqref{J_e-hd-hcme}]. Consequently, the peaks in  $\mathscr{C}_N(i,m,t)$ {\it vs.} $t$ in fig.~\ref{C-K-R=1}a also change sign. The peaks with opposite signs in fig.~\ref{C-K-R=1}a gives rise to rising, saturating and falling  structure of the time integrated correlation $\mathscr{K}_N(i,m,t)$ with time, as shown in fig.~\ref{C-K-R=1}b. The cumulative correlation for a given location $\alpha$ rises when a positive peak passes by this location and it falls when a negative peak passes by. The saturation appears when no peaks are passing by. Also note that cumulative correlation for different values of $\alpha$ finally approaches to a saturation value independent of $\alpha$ for system of fixed size in the large $t$ limit as they should by Eq.~\eqref{finite-N-kernel}.

Since for fixed BC case $R=1$, the kernel in Eq.~\eqref{kernel-hcme} has the form
\begin{align}
\mathcal{K}(u,v) = \frac{\omega^{3/2}}{2\sqrt{4\pi}} \Bigg{[}\underbrace{\frac{1}{\sqrt{v-u}}}_{0^{\text{th}}} - \underbrace{\frac{1}{\sqrt{v+u}}}_{1^{\text{st}}~\text{from left}} - \underbrace{\frac{1}{\sqrt{2-u-v}}}_{1^{\text{st}}~\text{from right}} + ...\Bigg{]},~\text{for}~~u <v. \label{mcalK_R-1}
\end{align}
The first term inside the bracket represents contribution from the event when the left moving sound peak starting at $v$ passes through $u$ before getting any reflections from either of the boundaries. The second term represents contribution from the events when the original left moving sound peak passes through $u$ again after getting reflected back from the left boundary. Similarly, the third term represents contribution from the events when the original right moving sound peak passes through $u$  after getting reflected back from the right boundary.

\begin{figure}
\begin{center}
\leavevmode
\includegraphics[scale=0.4]{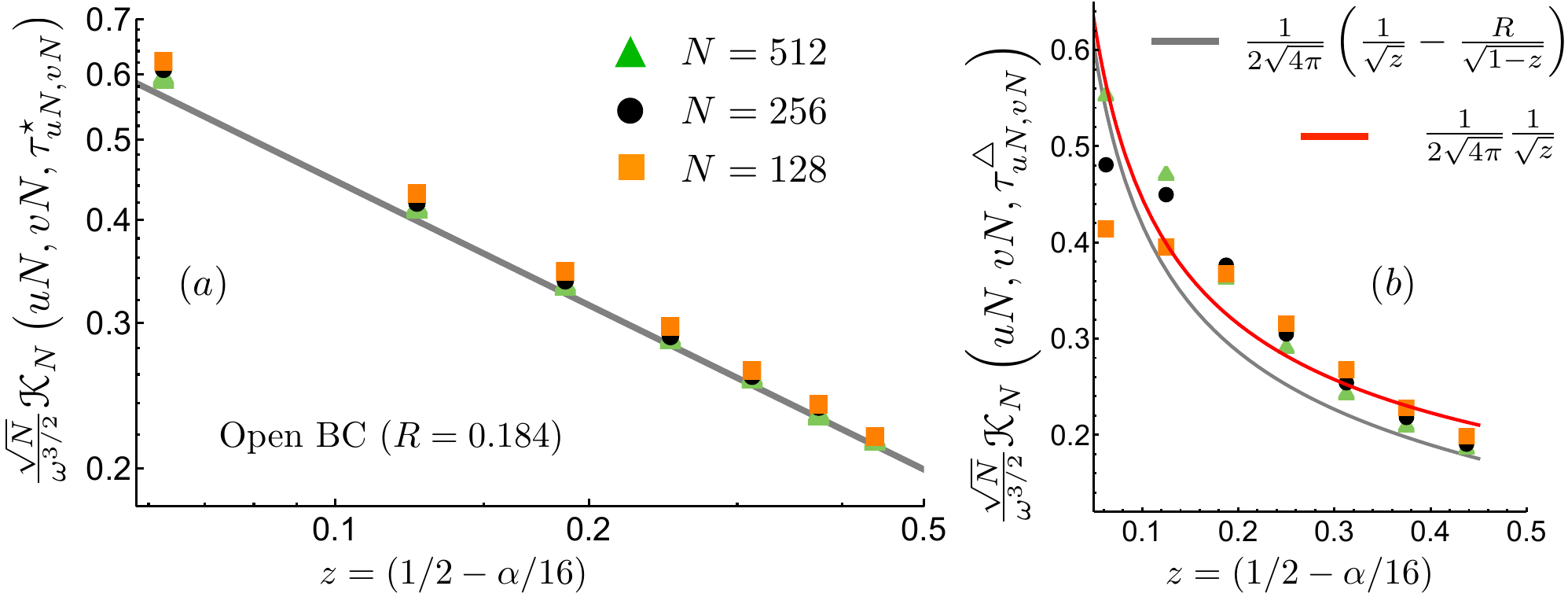}
%	\put (-150,120) {$\textbf{{\small(a)}}$}
%\includegraphics[scale=0.25]{fig/Ker-verify-R<1-2nd.pdf}	
%\includegraphics[width=7cm,angle=0]{figs/H-function-diff-dv-N50000_Scipost.eps}
%	\put (-30,120) {$\textbf{{\small(b)}}$}
\caption{Plots of (a) $\frac{\sqrt{N}}{\omega^{3/2}}\mathscr{K}_N\left(uN,vN,t^\star_{uN,vN}\right)$ {\it vs.} $z=v-u$  (b) $\frac{\sqrt{N}}{\omega^{3/2}}\mathscr{K}_N\left(uN,vN,t^\triangle_{uN,vN}\right)$ {\it vs.} $z$ with $v =\frac{1}{2}$ and $u=\frac{\alpha}{16}$ for $\alpha=1,2,...,7$. The solid gray lines in (a) and  in (b) represents the kernel till the first term  and the kernel till the second term on the right hand side (RHS) of Eq.~\eqref{mcalK_R<1}, respectively. The symbols corresponding to different $N$ are obtained from simulation. In figure (b) we have also plotted the first term $\frac{1}{2\sqrt{4\pi}}\frac{1}{\sqrt{z}}$ (solid red line) for reference. We have used $\lambda=1$ and $\omega=2.5$ for this plot. }
\label{num-verify-K-R<1-12}
\end{center}
\end{figure}

The $\star$ points in fig.~\ref{C-K-R=1}b represents the value of the cumulative correlation at times $t^\star$  when a peak passes through  the location $i=\frac{\alpha N}{16}$ for $\alpha=1,2,...,7$ for the first time (before getting any reflections from the boundaries). The  procedure for precisely estimating  the time $t^\star$ for different $\alpha$ was discussed in the $R=0$ case previously. In fig.~\ref{num-verify-K-R=1-12}a we demonstrate that the values of the scaled cumulative correlation at $t^\star$ for different $\alpha$ verifies the first term on the RHS of  Eq.~\eqref{mcalK_R-1}. In this figure we plot 
$\frac{\sqrt{N}}{\omega^{3/2}}\mathscr{K}_N\left(uN,vN,t^\star_{uN,vN}\right)$ as functions of $z=v-u$ in log-log scale for different $N$ with $v=\frac{1}{2}$ and $u=\frac{\alpha}{16}$ for $\alpha=1,2,...,7$. We observe very good agreement with the theory for increasing $N$.

Beyond time $t^\star$ the (left moving) sound peak gets reflected from the left boundary and gets converted to a right moving sound mode. This reflected sound mode now again passes through the location  $i=\frac{\alpha N}{16}$ at times represented by the x-coordinates $t^\triangle$ of the $\triangle$ points in fig.~\ref{C-K-R=1}b. In fig.~\ref{num-verify-K-R=1-12}b we plot the values of the scaled cumulative correlation $\frac{\sqrt{N}}{\omega^{3/2}}\mathscr{K}_N\left(uN,vN,t^\triangle_{uN,vN}\right)$ at the  $\triangle$ points as functions of $z=v-u$ for different $N$ with $v=\frac{1}{2}$ and $u=\frac{\alpha}{16}$ for $\alpha=1,2,...,7$. We compare the data with the theoretical expression 
$\frac{1}{2\sqrt{4\pi}} \left[\frac{1}{\sqrt{v-u}} - \frac{1}{\sqrt{v+u}} \right]$ obtained from the  the first two terms on the RHS of Eq.~\eqref{mcalK_R-1}. The agreement between the theory and simulation data verifies the analytical expression of the kernel $\mathcal{K}(u,v)$ in Eq.~\eqref{mcalK_R-1} up to the second term inside the bracket.  At smaller $z$, numerical data are still away from the theoretical curve possibly because for small $z$ the time difference between two successive crossing events is small and for smaller $N$ this time difference is below the hydrodynamic time scales. As a consequence the data at smaller $z$ converges slowly with increasing $N$.

Verifying higher order terms in the series associated to  reflections at later times seems difficult because of the following reasons. First, as time progresses the integrated correlation $\mathscr{K}_N(i,m,t)$ starts getting saturated to the final value because integration duration approaches infinity while the system size is kept fixed.
%at later times it becomes difficult to maintain the order of limits in Eq.~\eqref{def:mcalK} for a system with given size.  
Second, at later time the value of correlation $\mathscr{C}_N(i,m,t)$ itself becomes very small and possibly becomes comparable to statistical errors. Third, the peaks at later times get so broad due to diffusion that it becomes difficult to separate  the completion of one passing by event by a sound mode from the starting of the next  passing by event.

\begin{figure}
\begin{center}
\leavevmode
\includegraphics[scale=0.25]{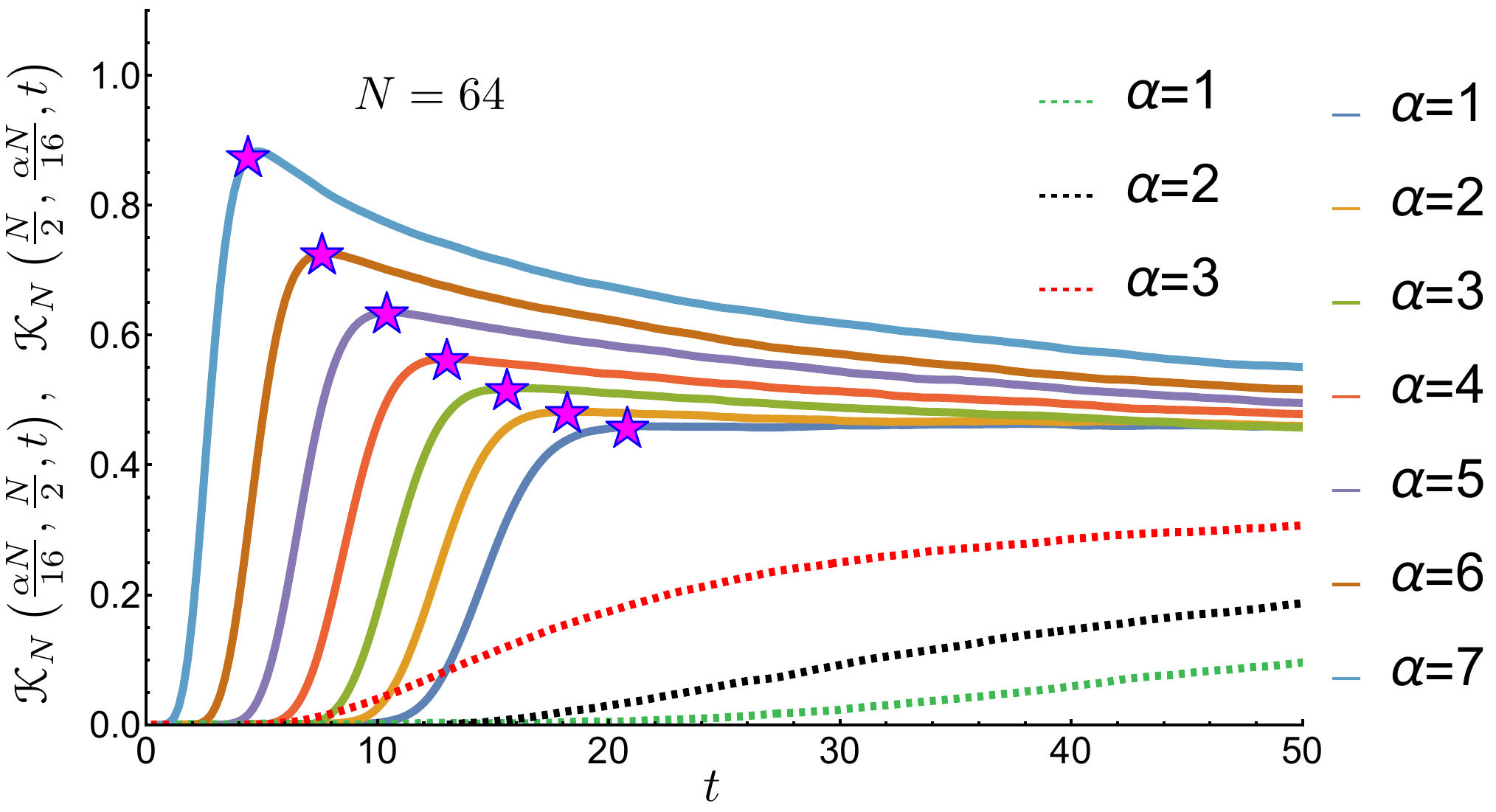}
%\includegraphics[scale=0.4]{fig/C-and-K-N64-ve.pdf}
%	\put (-170,120) {$\textbf{{\small(a)}}$}
%\includegraphics[width=7cm,angle=0]{figs/H-function-diff-dv-N50000_Scipost.eps}
%	\put (-170,120) {$\textbf{{\small(b)}}$}
\caption{Plots showing time evolutions of $\mathscr{K}_N(i,m,t)$ (solid lines) and $\mathscr{K}_N(m,i,t)$ (dotted lines)  for the HCVE model  with $m = \frac{N}{2}$ and $i=\frac{\alpha N}{16}$ where $\alpha=1,2,...,7$. Note that the integrated correlations $\mathscr{K}_N(i,m,t)$ ($i>m$ case) [dashed lines]
are smaller than $\mathscr{K}_N(i,m,t)$ ($i<m$ case) and approaches the saturation value exponentially. Parameters for this plot are $\lambda=1$ and $\omega=1$.  }
\label{C-K-ve}
\end{center}
\end{figure}

\subsubsection{$0<R<1$ case:} This case is similar to the $R=1$ case with the important difference is that the sound modes upon reflections lose some of its amplitudes at the boundaries. In the purely absorbing case ($R=0$) the sound modes get completely absorbed once they hit the boundaries, whereas for $0<R<1$ case they get partially absorbed. The first few terms of the kernel $\mathcal{K}(u,v)$ in this case are 
\begin{align}
\mathcal{K}(u,v) = \frac{\omega^{3/2}}{2\sqrt{4\pi}} \Bigg{[}\underbrace{\frac{1}{\sqrt{v-u}}}_{0^{\text{th}}} - \underbrace{\frac{R}{\sqrt{v+u}}}_{1^{\text{st}}~\text{from left}} - \underbrace{\frac{R}{\sqrt{2-u-v}}}_{1^{\text{st}}~\text{from right}} + ...\Bigg{]},~\text{for}~~u <v. \label{mcalK_R<1}
\end{align}
We provide numerical results corresponding to the first two terms as done for the $R=1$ case. To verify the first term, in fig.~\ref{num-verify-K-R<1-12}a we plot the values of the scaled cumulative correlation $\frac{\sqrt{N}}{\omega^{3/2}}\mathscr{K}_N\left(uN,vN,t^\star_{uN,vN}\right)$ at $t^\star$ times at which the left moving sound mode  crosses the location $u$ starting from $v=\frac{1}{2}$ for different values of $u=\alpha/16$ for $\alpha=1,2,...,7$. In the log-log scale we once again observe nice agreement with the analytical expression provided by the first term on the RHS of Eq.~\eqref{mcalK_R<1}. To verify the kernel up to second term, in fig.~\ref{num-verify-K-R<1-12}b we plot  $\frac{\sqrt{N}}{\omega^{3/2}}\mathscr{K}_N\left(uN,vN,t^\triangle_{uN,vN}\right)$ at $t^\triangle$ times when the initially left moving sound mode once again crosses the location $u$ after getting partially reflected back from the left boundary. We compare our data with analytical result coming from the first two terms of the kernel in Eq.~\eqref{mcalK_R<1} and observe that the numerical results approach the theoretical curve with increasing $N$ for larger $z$ values. At smaller $z$, numerical data are still away from theory, once again possibly because finite size effect is stronger at small $z$ after the first reflection.

\begin{figure}
\begin{center}
\leavevmode
\includegraphics[scale=0.3]{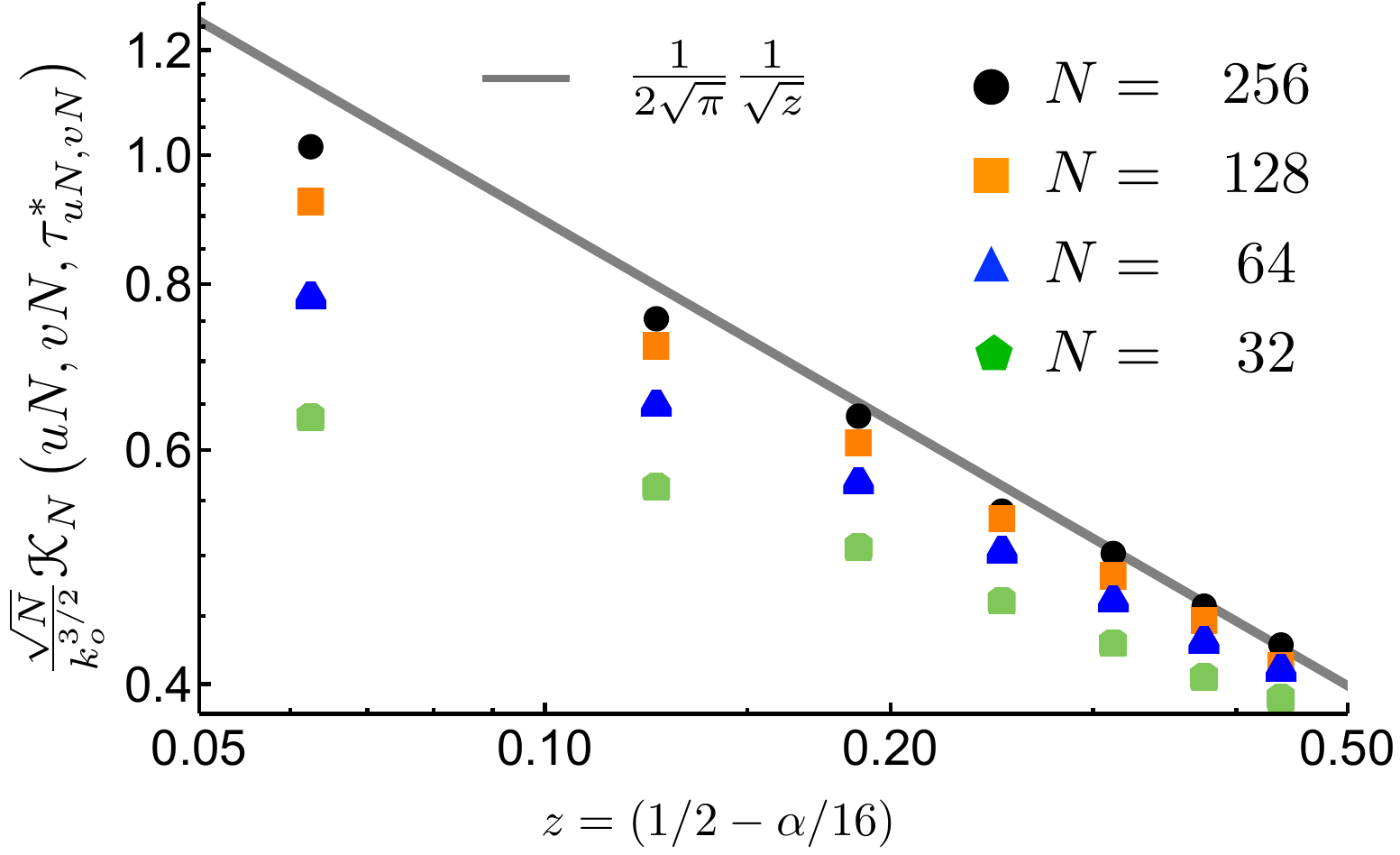}
%	\put (-150,120) {$\textbf{{\small(a)}}$}
%\includegraphics[scale=0.25]{fig/Ker-verify-R-1-2nd.pdf}	
%\includegraphics[width=7cm,angle=0]{figs/H-function-diff-dv-N50000_Scipost.eps}
%	\put (-30,120) {$\textbf{{\small(b)}}$}
\caption{Plots of $\frac{\sqrt{N}}{\omega^{3/2}}\mathscr{K}_N\left(uN,vN,t^\star_{uN,vN}\right)$ {\it vs.} $z=v-u$  with $v =\frac{1}{2}$ and $u=\frac{\alpha}{16}$ for $\alpha=1,2,...,7$. The solid grey line represents the kernel given in Eq.~\eqref{kernel-hcve}. The symbols corresponding to different $N$ are obtained from simulation.  We have used $k_{\rm o}=1$, $\lambda=1$ and $T=3$ for this plot.}
\label{num-verify-K-ve}
\end{center}
\end{figure}

\subsection{Results for HCVE model:}
Unlike the HCME model this system has two conserved quantities. Hence, in addition to the heat mode this model has only one sound mode which moves in the left direction with speed $c=2k_{\rm o}$. As a result the integrated correlation $\mathscr{K}_N(i,m,t)$ behaves differently for $i<m$ and $i>m$. In the former case, the currents at locations $i$ and $m$ get more correlated because the sound mode, starting from position $m$ crosses the position $i$ after time $t_{im}=\frac{m-i}{2k_{\rm o}}$. On the other hand for $i>m$, the currents at these two locations can  get correlated only through diffusive spreading of the sound mode as well as through bare diffusion of the heat mode itself. This can be observed in fig.~\ref{C-K-ve} where we plot $\mathscr{K}_N(i,m,t)$ as functions of time for different choices of $i$ and $m$. This asymmetric behavior in space gets manifested in the expression of the kernel in Eq.~\eqref{kernel-hcve} where one has a Heaviside theta function. To verify this expression numerically we follow the same procedure as done for the HCME case.  In fig.~\ref{num-verify-K-ve} we plot the values of the scaled cumulative correlation $\frac{\sqrt{N}}{k_{\rm o}^{3/2}}\mathscr{K}_N\left(uN,vN,t^\star_{uN,vN}\right)$ at $t^\star$ times at which the sound mode, starting from $v=\frac{1}{2}$,  crosses the location $u=\frac{\alpha}{16}$ with $\alpha=1,2,...,7$. As a function of $z=v-u$, these values indeed decay as given in Eq.~\eqref{kernel-hcve} [solid grey line]. The increasingly better agreement with increasing system size $N$ verifies the theoretical result.

\section{Conclusion}
\label{conclusion}
In this paper we addressed the important question on how non-local LR relation appears in anomalous transport through an open system in the HD limit from the space-time correlation of local microscopic currents. In systems with finite number of degrees of freedom we showed that the open system GK formula can not provide a space dependent kernel to characterise the non-local LR relation. On the other hand a LR theory based on HD currents provides a non-local LR relation in case of anomalous transport. We showed how starting from an open system GK formula involving microscopic currents measured at different locations of a system, one can achieve the non-local LR relation  in the appropriate combination of the limits of -- system size going to infinity and the  integral time duration (in the GK formula)  going to infinity limit. In particular, by computing time correlations of microscopic currents at different locations numerically, we  demonstrated the procedure to compute the detailed analytical form of the kernel operator characterising the non-local LR relation in the context of two microscopic models of anomalous transport, namely the HCME and HCVE systems.

The kernel $\mathcal{K}(u,v)$  governs the super-diffusion of the energy density field in an infinite isolated system. On the other hand, the kernel along with the boundary conditions also decides the NESS temperature profiles in an open system \cite{cividini2017temperature}. In case of normal transport the Fourier's law along with the continuity equation and local equation of state give rise to the diffusion equation for the evolution of temperature profile. Similarly, in the context of anomalous transport the generalisation of the Fourier's law as expressed by the non-local linear response relation also generalises the diffusion equation to a non-local (fractional) diffusion equation for studying the evolution of temperature profiles \cite{kundu2018anomalous}. Hence the detailed knowledge of the kernel operators in systems exhibiting anomalous transport  is important. Our study reveals, where, in the time integral of the space-time correlations of local currents, the information about the kernel operator is hiding and how to find it.

It would be interesting to obtain such kernel operators in other systems exhibiting anomalous transport, especially Hamiltonian systems like Fermi-Pasta-Ulam-Tsingou model. Secondly, since the operator is non-local it's spectral properties and Green's function will be highly sensitive to boundary conditions unlike the Laplacian operator \cite{kundu2019fractional}. It would also be interesting to study mathematical properties of such kernel operators.

\section{Acknowledgement}
The author would like to thank David Mukamel for the discussion from which this project got originated and very useful comments on the manuscript. The author also  acknowledges the support of the core research grant no. CRG/2021/002455 and MATRICS grant MTR/2021/000350 from the Science and Engineering Research Board (SERB), Department of Science and Technology, Government of India. A.K. also acknowledges support from the Department of Atomic Energy, Government of India, under project no. 19P1112R\&D.

\appendix
\section{Explicit expressions of the FP operators}
\label{app1}
For HCME model the FP operators are given by 
\begin{align}
\mathcal{L}_{\ell}P(\vec{\mu},t) &= -\sum_{i=1}^N\left[ p_i \partial_{q_i}  + \omega^2(q_{i+1}-2q_i+q_{i-1})\partial_{p_i}\right] P(\vec{\mu},t), \label{app:mcalL-hcme-l}\\
\mathcal{L}_{ex}P(\vec{\mu},t) &= \sum_{i=1}^{N-1}[P(\vec{\mu}_{i,i+1},t) -P(\vec{\mu},t) ],  \label{app:mcalL-hcme-ex}\\
\mathcal{L}_{b}P(\vec{\mu},t) &= \left[\lambda \partial_{ p_1} p_1+ \lambda T_L \partial_{p_1}^2 
+ \lambda  \partial_{p_N}p_N+ \lambda T_R \partial_{p_N}^2
\right] P(\vec{\mu},t), \label{app:mcalL-hcme-b}
\end{align}
where $\vec{\mu}=(q_1,q_2,...,q_N, p_1,p_2,...,p_N)$ and $\vec{\mu}_{i,i+1}$ is the state  after exchanging the momenta $p_i$ and $p_{i+1}$ in $\vec{\mu}$. 

For HCVE model the FP operators are given by 
\begin{align}
\mathcal{L}_{\ell}P(\vec{\eta},t)&= -k_{\rm o}\sum_{i=1}^N\left[\eta_{i+1} - \eta_{i-1}\right] \partial_{\eta_i}P(\vec{\eta},t), 
\label{app:mcalL-hcve-l}\\
\mathcal{L}_{ex}P(\vec{\eta},t)&= \sum_{i=1}^N\left[ P(\vec{\eta}_{i,i+1},t) - P(\vec{\eta},t) \right], \label{app:mcalL-hcve-b}\\
\mathcal{L}_{b}P(\vec{\eta},t) &= \left[
 \lambda k_{\rm o} \partial_{\eta_1} \eta_1  +\lambda T_L \partial_{\eta_1}^2  
 + \lambda k_{\rm o}\partial_{\eta_N} \eta_N  + \lambda T_R \partial_{\eta_N}^2 \right]P(\vec{\eta},t),
\label{app:mcalL-hcve-ex}
\end{align}
where $\vec{\eta}=(\eta_1,\eta_2,...,\eta_N)$ and $\vec{\eta}_{i,i+1}$ represents the state after exchanging $\eta_i$ and $\eta_{i+1}$ in $\vec{\eta}$.

\section{Explicit expressions of $\Phi$ and $\Phi_{le}$ defined in Eq.~\eqref{Phi}}
\label{app2}
Here we present the expressions of $\Phi$ and $\Phi_{le}$ for HCME model. A similar calculation can be carried out for the HCVE model following the same steps. Inserting the explicit form of the LE distribution from Eq.~\eqref{P_le-hcme} in Eq.~\eqref{Phi} and performing some algebraic simplifications we get 
\begin{align}
\Phi(\vec{\mu},t) &= \sum_{k=1}^N(\tau_k-\tau_{k-1})j^{(s)}_{k,k-1} + \sum_{k=1}^{N-1}(\pi_{k+1}-\pi_k)[j^{(p)}_{k+1,k}+(\sigma_{k+1,k}-\gamma) (p_{k}-p_{k+1})] \\
& +\sum_{k=1}^{N-1}(\beta_{k+1}-\beta_k) \left[j^{(e)}_{k+1,k}+(\sigma_{k+1,k}-\gamma) \left(\frac{p_{k}^2}{2}-\frac{p_{k+1}^2}{2}\right)\right] \\
&+(\beta_1-\beta_L)\frac{\lambda}{\beta_L}(\beta_1p_1^2-1) +(\beta_R-\beta_N)\frac{\lambda}{\beta_R}(1-\beta_Np_N^2) \\
&+\frac{\lambda \pi_1}{\beta_L} (2\beta_1p_1+\pi_1-\beta_Lp_1) +  \frac{\lambda \pi_N}{\beta_R} (2\beta_Np_N+\pi_N-\beta_Rp_N), 
\end{align}
where $\beta_i=1/T_i$ and $\beta_{L,R}=1/T_{L,R}$ and the expressions of the currents $j^{(s,p,e)}_{k+1,k}$ are given in Eq.~\eqref{hcme-micro-j}. The terms in the $3^{\text{rd}}$ and  $4^{\text{th}}$ lines are boundary terms. The expression of $\Phi_{le}(\vec{\mu},t)=\partial_t \ln P_{\rm le}(\vec{\mu},t)$ is given by 
\begin{align}
\Phi_{le}(\vec{\mu},t) \simeq -\sum_{k=1}^N[(e_k-e_{eq}) (\partial_t \beta_k)_{le} + (p_k-p_{eq}) (\partial_t \pi_k)_{le} +(s_k-s_{eq}) (\partial_t \tau_k)_{le}], 
%+ \text{terms independent of}~\vec{\mu}, 
\label{app:Phi_le}
\end{align} 
where $(...)_{eq}$ represents  average values in the underlying global equilibrium and $(\partial_t ...)_{le}$ represents the rate of change of the fields under the local equilibrium approximation in the LR regime. To find out these time derivatives we take average on both sides of the continuity equations in the LE state. We assume the LE state is slightly deviated from an underlying homogeneous global equilibrium (GE) state characterized by 
$\beta_0$, $\pi_0=0$ and $\tau_0 =0$. Hence the LE state is characterized by fields $\beta_k=\beta_0+\delta \beta_k,~\pi_k=\delta \pi_k$ and $\tau_k=\delta \tau_k$ such that the deviations are small.  Up to linear order in these deviations, one can easily show that  
\begin{align}
\langle s_k\rangle_k &\simeq \frac{1}{\omega^2} \frac{\delta \tau_k}{\beta_0}, ~~
\langle p_k\rangle_k \simeq \frac{\delta \pi_k}{\beta_0}, ~~
\langle e_k \rangle  \simeq \frac{1}{\beta_0} - \frac{\delta \beta_k}{\beta_0^2}.
\end{align}
Similarly the average of the currents in LE state can be computed to linear order in deviations. Inserting these averages in the continuity equations one gets
\begin{align}
(\partial_t \delta \tau_k)_{le} &= \omega^2 (\delta \pi_{k+1} -\delta \pi_k), \\
(\partial_t \delta \pi_k)_{le} &= (\delta \tau_{k} -\delta \tau_{k-1}) +\frac{\gamma}{\beta_0}\Delta_k \delta \pi_k, \\
(\partial_t \delta \beta_k)_{le} &= \gamma \Delta_k\delta \beta_k, 
\end{align}
where $\Delta_k$ is the discrete Laplacian.
Using these expressions in Eq.~\eqref{app:Phi_le} one gets $\Phi_{le}$, subtracting which from $\Phi$ and also taking average over the exchange noises, one finally gets 
\begin{align}
\begin{split}
\Phi-\Phi_{le} =
%&= \sum_{k=1}^{N-1} (\pi_{k}-\pi_{k}) (\sigma_{k+1,k}-\gamma) (p_{k+1}-p_k)  \\ 
&\sum_{k=1}^{N-1} (\beta_{k+1}-\beta_k)~j^{(e)}_{k+1,k} + \text{boundary terms + higher order derivative terms}.
\end{split}
\label{app:Phi-Phi_le}
\end{align}
%\section{Details of the derivation of the LR relation \eqref{LRR-1}}
%\label{app3}

%\clearpage
\section*{References}

\bibliographystyle{unsrt}
\bibliography{hcve}

\end{document}